\documentclass[aip,apl
 amsmath,amssymb,
 reprint,%
]{revtex4-1}

\usepackage{graphicx}
\usepackage{epstopdf}
\usepackage{comment}
\usepackage{dcolumn}
\usepackage{bm}
\usepackage{amsmath}
\usepackage{amssymb}
\usepackage{braket}
\usepackage[mathlines]{lineno}
\usepackage{units}
\usepackage{upgreek}
\usepackage{chemformula}
\usepackage{orcidlink}
\renewcommand{\selectlanguage}[1]{}
\definecolor{NewBlue}{rgb}{0, 0, 0.41}
\definecolor{NewRed}{rgb}{0.6, 0.07, 0.07}

\hypersetup{colorlinks,
   linkcolor=NewBlue,
    citecolor=NewBlue,
    urlcolor=NewRed}
\newcommand*{\figref}[2][]{%
  Fig.~\hyperref[{fig:#2}]{%
    \ref*{fig:#2}%
    \ifx\\(#1)\\%
    \else
      (#1)%
    \fi
  }%
}

\begin{document}

\preprint{AIP/123-QED}

\title{Large-Range Tuning and Stabilization of the Optical Transition of Diamond Tin-Vacancy Centers by In-Situ Strain Control}
\author{Julia M. Brevoord \orcidlink{0000-0002-8801-9616}}
\affiliation{QuTech and Kavli Institute of Nanoscience$,$ Delft University of Technology$,$ Delft 2628 CJ$,$ Netherlands}

\author{Leonardo G. C. Wienhoven \orcidlink{0009-0009-7745-3765}} \affiliation{QuTech and Kavli Institute of Nanoscience$,$ Delft University of Technology$,$ Delft 2628 CJ$,$ Netherlands}

\author{Nina Codreanu \orcidlink{0009-0006-6646-8396}}
\affiliation{QuTech and Kavli Institute of Nanoscience$,$ Delft University of Technology$,$ Delft 2628 CJ$,$ Netherlands}

\author{Tetsuro Ishiguro}
\affiliation{QuTech and Kavli Institute of Nanoscience$,$ Delft University of Technology$,$ Delft 2628 CJ$,$ Netherlands}
\affiliation{Quantum Laboratory, Fujitsu Limited, 10-1 Morinosato-Wakamiya, Atsugi, Kanagawa 243-0197, Japan}

\author{Elvis van Leeuwen}
\affiliation{QuTech and Kavli Institute of Nanoscience$,$ Delft University of Technology$,$ Delft 2628 CJ$,$ Netherlands}

\author{Mariagrazia Iuliano \orcidlink{0009-0003-4859-0521}}
\affiliation{QuTech and Kavli Institute of Nanoscience$,$ Delft University of Technology$,$ Delft 2628 CJ$,$ Netherlands}

\author{Lorenzo De Santis \orcidlink{0000-0003-0179-3412}}
\affiliation{QuTech and Kavli Institute of Nanoscience$,$ Delft University of Technology$,$ Delft 2628 CJ$,$ Netherlands}

\author{Christopher Waas \orcidlink{0009-0008-1878-2051}}
\affiliation{QuTech and Kavli Institute of Nanoscience$,$ Delft University of Technology$,$ Delft 2628 CJ$,$ Netherlands}

\author{Hans K. C. Beukers \orcidlink{0000-0001-9934-1099}}
\affiliation{QuTech and Kavli Institute of Nanoscience$,$ Delft University of Technology$,$ Delft 2628 CJ$,$ Netherlands}

\author{Tim Turan \orcidlink{0009-0003-9908-7985}}
\affiliation{QuTech and Kavli Institute of Nanoscience$,$ Delft University of Technology$,$ Delft 2628 CJ$,$ Netherlands}

\author{Carlos Errando-Herranz \orcidlink{0000-0001-7249-7392}}
\affiliation{QuTech and Kavli Institute of Nanoscience$,$ Delft University of Technology$,$ Delft 2628 CJ$,$ Netherlands}
\affiliation{Department of Quantum and Computer Engineering$,$ Delft University of Technology$,$ Delft 2628 CJ$,$ Netherlands}

\author{Kenichi Kawaguchi \orcidlink{0000-0002-9301-6321}}
\affiliation{Quantum Laboratory, Fujitsu Limited, 10-1 Morinosato-Wakamiya, Atsugi, Kanagawa 243-0197, Japan}

\author{Ronald Hanson \orcidlink{0000-0001-8938-2137}}
\affiliation{QuTech and Kavli Institute of Nanoscience$,$ Delft University of Technology$,$ Delft 2628 CJ$,$ Netherlands}
\email{R.Hanson@tudelft.nl}

\begin{abstract}
The negatively charged tin-vacancy (SnV$^-$) center in diamond has emerged as a promising platform for quantum computing and quantum networks. To connect SnV$^-$ qubits in large networks, in-situ tuning and stabilization of their optical transitions are essential to overcome static and dynamic frequency offsets induced by the local environment. Here we report on the large-range optical frequency tuning of diamond SnV$^-$ centers using micro-electro-mechanically mediated strain control in photonic integrated waveguide devices. We realize a tuning range of $>$\unit[40]{GHz}, covering a major part of the inhomogeneous distribution. In addition, we employ real-time feedback on the strain environment to stabilize the resonant frequency and mitigate spectral wandering. These results provide a path for on-chip scaling of diamond SnV-based quantum networks.
\end{abstract}

\maketitle
Spin qubits in the solid state are promising building blocks for the realization of large-scale quantum systems \cite{kimble_quantum_2008,wehner_quantum_2018,stas_robust_2022,nguyen_integrated_2019, pompili_realization_2021, stolk_metropolitan-scale_2024, li_heterogeneous_2024, bhaskar_experimental_2020}, with potential applications in computing and communication~\cite{kimble_quantum_2008, wehner_quantum_2018}. An early workhorse for the field has been the Nitrogen-Vacancy (NV) in diamond, which was used for a loophole-free Bell test~\cite{hensen_loophole-free_2015}, the realization of a multi-node quantum network~\cite{pompili_realization_2021} and heralded entanglement over metropolitan distances~\cite{stolk_metropolitan-scale_2024}. However, the relatively low Debye-Waller factor and the incompatibility with nanophotonic integration due to a first-order sensitivity to electrical noise limits the development of large-scale quantum networks and the integration into scalable photonic circuits based on NV-centers in diamond. Next-generation color centers in solid state materials have started to emerge over the last decade to drive scaling of quantum systems~\cite{bhaskar_experimental_2020,wan_large-scale_2020, heiler_spectral_2024,simmons_scalable_2024, bhaskar_quantum_2017}. The negatively charged tin-vacancy (SnV$^-$) center in diamond has attracted much recent attention due to its excellent optical properties~\cite{trusheim_transform-limited_2020, brevoord_heralded_2024,rugar_quantum_2021}, high quantum efficiency~\cite{iwasaki_tin-vacancy_2017, gorlitz_spectroscopic_2020}, and significant spin-orbit coupling~\cite{rosenthal_microwave_2023, karapatzakis_microwave_2024}, which allows for operation above dilution refrigerator temperatures~\cite{guo_microwave-based_2023, rosenthal_microwave_2023, beukers_control_2024}. Furthermore, the SnV$^-$ center lacks a first-order sensitivity to electrical noise due to its inversion symmetry, making it compatible with nanophotonic integration~\cite{rugar_characterization_2019,pasini_nonlinear_2024, clark_nanoelectromechanical_2024, li_heterogeneous_2024, arjona_martinez_photonic_2022, bradac_quantum_2019}. The recent demonstration of entanglement between the electron spin-$1/2$ SnV$^-$ center and a nearby $^{13}$C nuclear spin opens a path towards multi-qubit experiments using this platform~\cite{beukers_control_2024}. \\
\\
For connecting SnV$^-$ qubits into larger networks, photonic links mediating entanglement generation provide a modular and scalable path~\cite{ruf_quantum_2021, choi_percolation-based_2019, nickerson_freely_2014, nemoto_photonic_2014}. The entanglement generation requires the qubits to emit indistinguishable photons. However, the photon emission of spin qubits in solids is typically not indistinguishable due to different local strain in the material resulting in a broad inhomogeneous distribution. In addition, the optical resonance frequency variations over time due to changes in e.g. local strain and variations in the local electronic environment limit the indistinguishability of the emitted photons. This poses challenges for the realization of large-scale quantum networks. SnV$^-$ centers in diamond are first-order insensitive to electric fields, therefore conventional tuning using DC Stark tuning is limited~\cite{de_santis_investigation_2021, aghaeimeibodi_electrical_2021}. Previous work on strain engineering of other group-IV defects in diamond demonstrated frequency tuning~\cite{maity_spectral_2018, wan_large-scale_2020, meesala_strain_2018} and the mitigation of phonon interactions~\cite{sohn_controlling_2018}, and quantum interference of electromechanical stabilized emitters~\cite{machielse_quantum_2019}. Recent work on strain engineering of SnV$^-$ centers shows heterogeneous integration with efficient state preparation and readout~\cite{li_heterogeneous_2024, clark_nanoelectromechanical_2024} and a tuning range up to $\approx$~\unit[25]{GHz}. Here, we integrated diamond waveguide devices containing SnV$^-$ centers with local optical frequency control via electromechanically induced strain. We demonstrate a tuning range of $>$\unit[40]{GHz} and use the strain control in a feedback loop realizing a 12-fold improvement in the stabilization of the optical transition.\\
\\
The SnV$^-$ center in diamond consists of a tin-atom at the interstitial position between two missing carbon atoms with a captured electron, as shown in~\figref[a]{fig1}. A simplified level structure of the SnV$^-$ center is given in~\figref[b]{fig1}. In the absence of a magnetic field, the ground and optically excited states are composed of spin-degenerate orbital doublets, resulting in four optical transitions~\cite{hepp_electronic_2014, iwasaki_tin-vacancy_2017}. In the presence of strain, the energy levels shift. In this work, we control the optical transition linking the lowest branch of the ground and the optically excited state (ZPL), the C-transition. The C-transition is well suited for spin readout and photon emission for photon-mediated entanglement generation. We study SnV$^-$ centers in our device in an optical confocal set-up at an operating temperature of \unit[4]{K}, see Supplementary Material for more details on the experimental set-up. The variation in optical resonances between different color centers in the diamond sample (inhomogeneous distribution) will determine the tuning range required to produce indistinguishable photons from multiple color centers within the same diamond sample. The inhomogeneous distribution of the SnV$^-$ centers in the devices fabricated on a diamond sample used for this work is determined by photoluminescence (PL) measurements. We collect the emission in free space using a spectrometer while illuminating the sample with an off-resonant \unit[515]{nm} laser with \unit[200]{$\upmu$W} and integrating over \unit[10]{s} at 173 different laser spot locations on different devices. We determine the resonances of the C-transition using a peak-finding algorithm. The cumulative inhomogeneous distribution is plotted in~\figref[c]{fig1}. About 40$\%$ of the resonances detected in the region of interest are within a \unit[40]{GHz} window. This range is not a fundamental limitation, in the recent work of Li et al. ~\cite{li_heterogeneous_2024}, they found 40$\%$ of the resonances in a $\approx$~\unit[20]{GHz} window. Moreover, Görlitz et al. reported in~\cite{gorlitz_spectroscopic_2020} a linewidth of \unit[15]{GHz} for the ZPL transition of an ensemble of SnV centers in a high-pressure, high-temperature annealed diamond implanted with Sn atoms. Besides overcoming this static inhomogeneity, spectral diffusion and bi-stability pose a second challenge, as reported by others~\cite{pasini_nonlinear_2024, brevoord_heralded_2024,arjona_martinez_photonic_2022, pieplow_quantum_2024, herrmann_coherent_2024, li_atomic_2024}. In this work, we target both challenges using in-situ strain engineering of the SnV$^-$ including real-time feedback.
\begin{figure}
	\includegraphics[width=\linewidth]{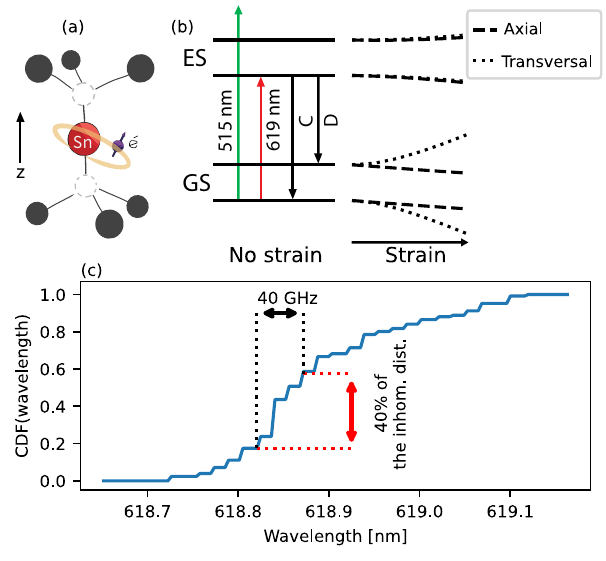}
	\caption{(a) Schematic of the SnV$^-$ center structure. (b) The energy level scheme of an SnV$^-$ center in diamond. We focus on tuning and stabilizing the C-transition, which is the most relevant for quantum technology applications. Resonant excitation at \unit[619]{nm} results in zero-phonon emission via the C-transition and the D-transition, as well as their respective phonon side bands (PSB, not shown). Strain alters the energy levels. (c) The cumulative distribution function (CDF) of the inhomogeneous distribution from resonances obtained with PL. }\label{fig:fig1}
\end{figure}
\\
\\
Following Meesala, Sohn et al. \cite{meesala_strain_2018} and Guo, Stramma, et al. \cite{guo_microwave-based_2023}, considering an infinitesimal strain regime, we use the Hamiltonian describing strain in the basis of the SnV$^-$:
\begin{equation}
    \mathbb{H}_{\text{strain}} = \sum_{i,j}A_{ij}\epsilon_{ij},
\end{equation}
where $\epsilon_{ij}$ is a component of the strain tensor, $\bm{\epsilon}$, and $A_{ij}$ operators that act on the electronic levels. Using group theory, we can write the Hamiltonian in the irreducible representations of the $D_{3d}$ point group, in the basis $\{ |e_x\uparrow\rangle$, $|e_x \downarrow\rangle$,$|e_y \uparrow\rangle$,$|e_y \downarrow\rangle$\}, where $e_x$, $e_y$ represent the orbital and $\downarrow$, $\uparrow$ the electronic spin-1/2 states. This shows that strain does not couple the ground and excited states, resulting in identical strain Hamiltonians forms for the ground and excited states. In this basis and in the local coordinate frame of the SnV$^-$ center, where z is along the high-symmetry axis, the $[$111$]$~crystallographic direction, the part of the Hamiltonian describing the presence of strain is,
\begin{equation}
    \mathbb{H}_{\text{strain}} = \begin{bmatrix}
\epsilon_{A_{1g}} - \epsilon_{E_{gx}} & \epsilon_{E_{gy}} \\
\epsilon_{E_{gy}} & \epsilon_{A_{1g}} + \epsilon_{E_{gx}}
\end{bmatrix}  \otimes \mathbb{I}_2 ,
\end{equation}
where $\epsilon_{A_{1g}}, \epsilon_{E_{gx}}, \epsilon_{E_{gy}}$ represent the strain induced deformation modes in the $D_{3d}$ point group. Rewriting each term as a linear combination of the strain tensor results in,
\begin{equation}
\begin{aligned}
    \epsilon_{A_{1g}} &= t_{\perp}(\epsilon_{xx}+\epsilon_{yy})+ t_{\parallel}\epsilon_{zz}, \\
    \epsilon_{E_{gx}} &= d(\epsilon_{xx}-\epsilon_{yy})+ f\epsilon_{zx}, \\
    \epsilon_{E_{gy}} &= -2d \epsilon_{xy}+ f\epsilon_{yz},
    \end{aligned}
    \label{eq:strain energy levels}
\end{equation}
where $t_{\perp},t_{\parallel},d$ and $f$ are the strain susceptibilities, with different values in the ground and excited state. Diagonalizing the Hamiltonian and including spin-orbit coupling results in expressions for the mean ZPL frequency,  $\nu_{\text{ZPL}}$, and ground and excited state splitting, $\Delta_{u,g}$,
\begin{equation}
    \begin{aligned}
        \nu_{\text{ZPL}} &=\nu_0 + \epsilon_{A_{1g}}^u-\epsilon_{A_{1g}}^g, \\
        \Delta_u &= \sqrt{(\lambda_{SO}^u)^2+4(\epsilon_{E_{gx}}^u)^2+4(\epsilon_{E_{gy}}^u)^2}, \\
        \Delta_g &= \sqrt{(\lambda_{SO}^g)^2+4(\epsilon_{E_{gx}}^g)^2+4(\epsilon_{E_{gy}}^g)^2},
    \end{aligned}
    \label{eq:strain response}
\end{equation}
where $\nu_0$ is the mean ZPL frequency in the absence of strain and $\lambda_{SO}^{g, (u)}$ is the spin-orbit coupling of the ground (excited) state. The superscript $g$ ($u$) indicates the ground (excited) state contribution. \\

\begin{figure*}
	\includegraphics[width=\textwidth] {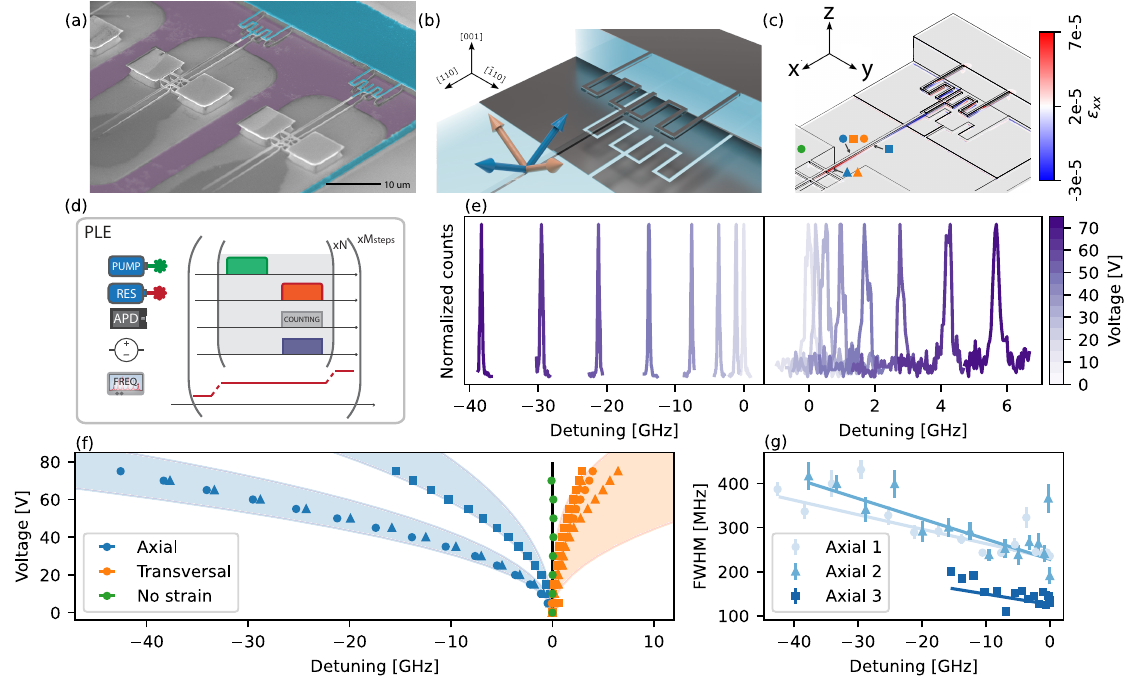}
	\caption{(a) False colored SEM image of the devices. (b) Orientation of the axial (transversal) SnV$^-$ centers in the waveguide, indicated by the blue (orange) arrows. (c) FEM simulation of the strain distribution ($\epsilon_{xx}$ component) over the surface of the device when \unit[75]{V} is applied. The estimated positions of the SnV$^-$ centers evaluated in (f) are indicated. (d) Pulse sequence of a photoluminescence excitation (PLE) scan. (e) Fluorescence of PLE scans taken at different bias voltages of an axial (blue triangle) (transversal (orange triangle)) SnV$^-$ center in the left (right) panel, showing an optical frequency shift of almost \unit[-40]{GHz} (\unit[6]{GHz}). The fluorescence is normalized for each PLE scan. (f) Summary of the measured optical resonance frequency detuning of several SnV$^-$ centers, indicated by different shaped data points, as a function of the applied bias voltage, the error bars lie within the data points. The shaded areas are the simulated frequency shift as a function of the applied bias voltage for the SnV$^-$ center \unit[20]{nm} below the surface at the hinge point (orange and blue triangle) and \unit[4]{$\upmu$m} away from it (blue square), the uncertainty is in the depth of the color center and the strain susceptibility parameter $t_{\perp}$. (g) The fitted FWHM as a function of detuning for multiple axial-oriented SnV$^-$ centers, and is fitted with a linear function. } \label{fig:fig2}
\end{figure*}
Our investigation focuses on SnV$^-$ centers embedded in waveguide-based micro-electro-mechanical system (MEMS) devices.  \figref[a]{fig2} shows a false-coloured scanning electron microscope (SEM) image of the full MEMS device. The SnV$^-$ incorporated waveguide is on one side connected via a mechanical spring-like structure to the bulk, and on the other side clamped by two bars to two diamond bulk support platforms. Niobium electrodes (with titanium as the adhesion layer) are deposited on the spring (top electrode), indicated by the blue-colored region in~\figref[a]{fig2}, and below the spring (bottom electrode), indicated by the purple shaded region on~\figref[a]{fig2}. These electrodes form a capacitive electromechanical actuator when a bias voltage is applied between the electrodes. According to Eq.~\ref{eq:strain response}, by varying the applied bias voltage we can tune the optical resonance frequency of the SnV$^-$ center due to the induced strain in the crystal. 

Our device is fabricated from an electronic grade $<$001$>$-surface-orientated diamond sample. The Sn-atoms are uniformly implanted with a dose of~\unit[10$^{11}$]{ions/cm$^2$}. After vacuum-annealing to activate the SnV$^-$ centers, the color centers are predicted to end up at a mean depth of \unit[90]{nm} below the surface with a straggle of \unit[17]{nm} from Stopping and Range of Ions in Matter (SRIM) simulations~\cite{ziegler_srim_2010}. Due to substantial graphite formation during the vacuum-annealing, an additional short plasma etching step was performed to remove the graphitic layer, resulting in SnV$^-$ centers closer to the surface. Fabrication details can be found in the Supplementary Material. The waveguides used in this work are longitudinally aligned along the [110] direction. The four possible orientations of SnV$^-$centers in the diamond crystal are [111], [$\overline{1}$11], [1$\overline{1}$1], and [$\overline{1}$$\overline{1}$1]. Considering the dominant uniaxial strain along the [110] direction when a voltage is applied, two groups of color centers that respond similarly to strain remain, indicated in~\figref[b]{fig2}. We denote the SnV$^-$ centers with a component of the dipole along (perpendicular) the long axis of the beam as axial (transversal) color centers. \\
\\
We perform a finite element method (FEM) simulation using Ansys~\cite{noauthor_ansys_nodate} to determine the strain in the waveguide when a voltage is applied over the electrodes. The $\epsilon_{xx}$ (lab frame) component of the strain tensor on the surface of the beam is shown in~\figref[c]{fig2} when \unit[75]{V} is applied, resulting in a maximal stain of $\epsilon_{xx} = $7$\times$10$^{-5}$. Additional information on the simulations can be found in the Supplementary Information. We have indicated the estimated position of the SnV$^-$ centers, positioned in different waveguides, evaluated in this work. \\
\\
By rotating the strain tensor obtained from the FEM simulation from the lab frame to the axial/transversal SnV$^-$ center reference frame and using Eq.~\ref{eq:strain energy levels} and Eq.~\ref{eq:strain response}, we simulate the resonant frequency shift of SnV$^-$ centers in our device as a function of applied voltage. We use the strain susceptibilities $d$ and $f$ determined by Guo and Stramma et al.~\cite{guo_microwave-based_2023} and $t_{\parallel}$ from studies on other group-IV color centers by Meesala and Sohn et al., \cite{meesala_strain_2018} and Maity and Shao et al.~\cite{maity_spectral_2018}. We note that using the single reported value for $t_{\parallel}$ for SnV$^-$ in~\cite{clark_nanoelectromechanical_2024} resulted in a discrepancy with the data well outside the uncertainty margin. For $t_{\perp}$ we take a lower and an upper bound based on reported values for other group-IV color centers~\cite{meesala_strain_2018,maity_spectral_2018}, and incorporate this into the uncertainty margins of our simulations.\\
\\
To determine the tuning range of the SnV$^-$ centers in this sample, we perform photoluminescence excitation (PLE) scans using the pulse sequence shown in~\figref[d]{fig2}. We record the phonon-side band (PSB) emission during the PLE scans. The bias voltage is applied in a pulsed way concurrent with the resonant readout to avoid heating originating from leakage current in the system. We have calibrated the pulse time and the time between subsequent pulses to eliminate the effects of heating. More information on the calibration can be found in the Supplementary Information.\\
\\
\figref[e]{fig2} shows the PSB collection during PLE scans for different applied bias voltages. We observe the two distinct tuning behaviors attributed to the axial and transversal-oriented groups. The results of multiple SnV$^-$ centers are summarized in \figref[f]{fig2}. For an axial and a transversal SnV$^-$ center positioned close to the hinge point (indicated by the orange and blue triangles in~\figref[c]{fig2}) and an axial SnV$^-$ center further away from the hinge point (indicated by the blue square), we simulate the expected frequency shift, indicated by the shaded areas in~\figref[e]{fig2}. We included an error bar representing our uncertainty in the depth of the color centers in the beam and the stain susceptibility values.  The optical resonance of an SnV$^-$ in the bulk (indicated by the green circle), which should not experience strain due to the bias voltage, is recorded for reference. This reference SnV$^-$ indeed shows no frequency shift with increasing voltage. We find that SnV$^-$ centers that show axial SnV$^-$ centers behavior can be tuned up to $\approx$~\unit[43]{GHz} and transversal centers up to $\approx$~\unit[6]{GHz}, covering a substantial part of the inhomogeneous distribution.

We observe a linear increase of the linewidth with frequency shift and hence strain in \figref[g]{fig2}, where we summarize the fitted FWHM of the linewidth obtained from the PLE scans of several axial SnV$^-$ centers. We observe a similar trend for these color centers,  with a mean linewidth increase of~\unit[3.42]{MHz/GHz}. Further investigation is needed to determine whether this behavior is intrinsic to SnV$^-$ centers or device-related (e.g. induced by local heating).\\
\\
The second challenge that we address in this work is the optical transition frequency drift over time. Drifts of the optical resonance frequency due to changes in the local charge environment can be compensated by real-time logic protocols, called Charge Resonance (CR) checks, that can herald the desired resonance condition~\cite{ bernien_heralded_2013,brevoord_heralded_2024}. However, drifts beyond the local-charge-induced distribution will lead to vanishing heralding success probabilities for these protocols. In those cases, the drift can be compensated for by dynamically adjusting the local strain. We implement a real-time feedback loop, utilizing strain tuning to stabilize the SnV$^-$ center at a desired resonant optical frequency. The experimental protocol is schematically depicted in~\figref[a]{fig3}. It starts with a resonant readout pulse at the target resonant frequency while counting the PSB photons. We superimpose the strain-inducing DC bias voltage with a sinusoidal signal and correlate the number of counts with the phase of the control modulation signal. This is followed by a CR check. Once this is passed, the experimental sequence is started. By averaging the photon counts obtained during the gate modulation phase of this sequence, we obtain an error signal to adjust the static DC bias voltage using a standard PID feedback loop. The strain-inducing DC bias voltage is updated at a~\unit[5]{Hz} rate. \\
\\
To test the stability we probe the resonant frequency of an SnV$^-$ for more than 7 hours by PLE scans. We fit the photon counts recorded per scan with a Voigt function and plot the fitted center frequency as a function of time in~\figref[b]{fig3} bottom panel. The stability observed over the 7-hour duration demonstrates the robustness of the system, and there are no intrinsic limitations that would preclude maintaining this stability over extended time periods. When no active feedback is applied, the drifts of the optical resonance are larger than can be compensated for by changes in the local charge environment. We compare the gate-modulated CR-checked PLE scans with PLE scans when no active feedback or CR checks are applied. The standard deviation of the fitted centers with active feedback is~\unit[30]{MHz}, while for the non-feedback case, the spread in fitted centers is \unit[1.38]{GHz}. In the top panel of ~\figref[b]{fig3} we plot a histogram of the fitted centers. The FWHM of the summed counts is~\unit[229]{MHz} $\pm$ \unit[2]{MHz} and the mean of the FWHM of the individual fitted scans is~\unit[225]{MHz}. Comparing the standard deviation in fitted centers with the non-feedback case results in a 12-fold increase in optical frequency stability with this feedback technique. 
\begin{figure}
	\includegraphics[width=\linewidth]{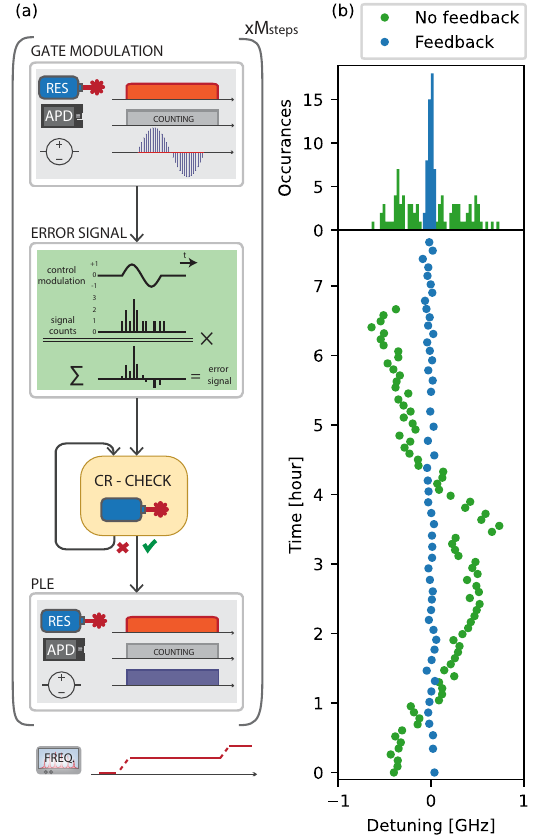}
	\caption{(a) Pulse sequence and real-time logic of the gate modulated CR-checked PLE. (b) (Bottom panel) Fitted centers of the fluorescence of 50 gate modulated CR-checked PLE scans, taken over 7 hours. For comparison the fitted centers of PLE scans with no active feedback in green. The error bar lies within the data points. (Top panel) Histogram of the fitted centers.} \label{fig:fig3}
\end{figure}
\\

In summary, we have presented and demonstrated fabricated devices that show a tunability up to $>$\unit[40]{GHz}, covering a significant part of the inhomogeneous distribution of the SnV$^-$ centers in this sample. We have stabilized the optical resonance of an SnV$^-$ using a dynamic strain feedback loop and demonstrated a 12-fold improvement in stability. The strain tuning is applied locally to SnV$^-$ embedded in waveguides and can be extended to tune numerous SnV$^-$ on the same chip simultaneously. The methods demonstrated here pave the way toward the generation of heralded indistinguishable photons from many integrated SnV$^-$ qubits, providing an important element for scaling SnV$^-$-based quantum technologies. \\
\\
The data that support the findings of this study are openly available in 4TU.ResearchData, at \url{https://www.doi.org/10.4121/6b89815e-d73c-4abd-93cc-a0514c2780ae}~\cite{brevoord_data_2024}.
\\

The authors want to thank R. Schouten and B. Otto for electronic support, N. Albers for machining parts of the experimental set-up, and O. Benningshof and J. Mensingh for cryogenics matter. The authors thank V. V. Dobrovitski for the fruitful discussions and A. Stramma for proofreading the manuscript. We acknowledge support from the joint research program “Modular quantum computers” by Fujitsu Limited and Delft University of Technology, co-funded by the Netherlands Enterprise Agency under project number PPS2007, from the Dutch Research Council (NWO) through the Spinoza prize 2019 (project number SPI 63-264), from the Dutch Ministry of Economic Affairs and Climate Policy (EZK) as part of the Quantum Delta NL program, from the Quantum Internet Alliance through the Horizon Europe program (grant agreement No. 101080128), from the European Union's Horizon Europe research and innovation program under grant agreement No. 101102140 – QIA Phase 1.
\bibliography{main.bib}

\begin{thebibliography}{53}%
\makeatletter
\providecommand \@ifxundefined [1]{%
 \@ifx{#1\undefined}
}%
\providecommand \@ifnum [1]{%
 \ifnum #1\expandafter \@firstoftwo
 \else \expandafter \@secondoftwo
 \fi
}%
\providecommand \@ifx [1]{%
 \ifx #1\expandafter \@firstoftwo
 \else \expandafter \@secondoftwo
 \fi
}%
\providecommand \natexlab [1]{#1}%
\providecommand \enquote  [1]{``#1''}%
\providecommand \bibnamefont  [1]{#1}%
\providecommand \bibfnamefont [1]{#1}%
\providecommand \citenamefont [1]{#1}%
\providecommand \href@noop [0]{\@secondoftwo}%
\providecommand \href [0]{\begingroup \@sanitize@url \@href}%
\providecommand \@href[1]{\@@startlink{#1}\@@href}%
\providecommand \@@href[1]{\endgroup#1\@@endlink}%
\providecommand \@sanitize@url [0]{\catcode `\\12\catcode `\$12\catcode `\&12\catcode `\#12\catcode `\^12\catcode `\_12\catcode `\%12\relax}%
\providecommand \@@startlink[1]{}%
\providecommand \@@endlink[0]{}%
\providecommand \url  [0]{\begingroup\@sanitize@url \@url }%
\providecommand \@url [1]{\endgroup\@href {#1}{\urlprefix }}%
\providecommand \urlprefix  [0]{URL }%
\providecommand \Eprint [0]{\href }%
\providecommand \doibase [0]{http://dx.doi.org/}%
\providecommand \selectlanguage [0]{\@gobble}%
\providecommand \bibinfo  [0]{\@secondoftwo}%
\providecommand \bibfield  [0]{\@secondoftwo}%
\providecommand \translation [1]{[#1]}%
\providecommand \BibitemOpen [0]{}%
\providecommand \bibitemStop [0]{}%
\providecommand \bibitemNoStop [0]{.\EOS\space}%
\providecommand \EOS [0]{\spacefactor3000\relax}%
\providecommand \BibitemShut  [1]{\csname bibitem#1\endcsname}%
\let\auto@bib@innerbib\@empty
\bibitem [{\citenamefont {Kimble}(2008)}]{kimble_quantum_2008}%
  \BibitemOpen
  \bibfield  {author} {\bibinfo {author} {\bibfnamefont {H.~J.}\ \bibnamefont {Kimble}},\ }\bibfield  {title} {\enquote {\bibinfo {title} {The quantum internet},}\ }\href {\doibase 10.1038/nature07127} {\bibfield  {journal} {\bibinfo  {journal} {Nature}\ }\textbf {\bibinfo {volume} {453}},\ \bibinfo {pages} {1023--1030} (\bibinfo {year} {2008})}\BibitemShut {NoStop}%
\bibitem [{\citenamefont {Wehner}, \citenamefont {Elkouss},\ and\ \citenamefont {Hanson}(2018)}]{wehner_quantum_2018}%
  \BibitemOpen
  \bibfield  {author} {\bibinfo {author} {\bibfnamefont {S.}~\bibnamefont {Wehner}}, \bibinfo {author} {\bibfnamefont {D.}~\bibnamefont {Elkouss}}, \ and\ \bibinfo {author} {\bibfnamefont {R.}~\bibnamefont {Hanson}},\ }\bibfield  {title} {\enquote {\bibinfo {title} {Quantum internet: {A} vision for the road ahead},}\ }\href {\doibase 10.1126/science.aam9288} {\bibfield  {journal} {\bibinfo  {journal} {Science}\ }\textbf {\bibinfo {volume} {362}} (\bibinfo {year} {2018}),\ 10.1126/science.aam9288}\BibitemShut {NoStop}%
\bibitem [{\citenamefont {Stas}\ \emph {et~al.}(2022)\citenamefont {Stas}, \citenamefont {Huan}, \citenamefont {Machielse}, \citenamefont {Knall}, \citenamefont {Suleymanzade}, \citenamefont {Pingault}, \citenamefont {Sutula}, \citenamefont {Ding}, \citenamefont {Knaut}, \citenamefont {Assumpcao}, \citenamefont {Wei}, \citenamefont {Bhaskar}, \citenamefont {Riedinger}, \citenamefont {Sukachev}, \citenamefont {Park}, \citenamefont {Lončar}, \citenamefont {Levonian},\ and\ \citenamefont {Lukin}}]{stas_robust_2022}%
  \BibitemOpen
  \bibfield  {author} {\bibinfo {author} {\bibfnamefont {P.-J.}\ \bibnamefont {Stas}}, \bibinfo {author} {\bibfnamefont {Y.~Q.}\ \bibnamefont {Huan}}, \bibinfo {author} {\bibfnamefont {B.}~\bibnamefont {Machielse}}, \bibinfo {author} {\bibfnamefont {E.~N.}\ \bibnamefont {Knall}}, \bibinfo {author} {\bibfnamefont {A.}~\bibnamefont {Suleymanzade}}, \bibinfo {author} {\bibfnamefont {B.}~\bibnamefont {Pingault}}, \bibinfo {author} {\bibfnamefont {M.}~\bibnamefont {Sutula}}, \bibinfo {author} {\bibfnamefont {S.~W.}\ \bibnamefont {Ding}}, \bibinfo {author} {\bibfnamefont {C.~M.}\ \bibnamefont {Knaut}}, \bibinfo {author} {\bibfnamefont {D.~R.}\ \bibnamefont {Assumpcao}}, \bibinfo {author} {\bibfnamefont {Y.-C.}\ \bibnamefont {Wei}}, \bibinfo {author} {\bibfnamefont {M.~K.}\ \bibnamefont {Bhaskar}}, \bibinfo {author} {\bibfnamefont {R.}~\bibnamefont {Riedinger}}, \bibinfo {author} {\bibfnamefont {D.~D.}\ \bibnamefont {Sukachev}}, \bibinfo {author} {\bibfnamefont {H.}~\bibnamefont {Park}}, \bibinfo {author}
  {\bibfnamefont {M.}~\bibnamefont {Lončar}}, \bibinfo {author} {\bibfnamefont {D.~S.}\ \bibnamefont {Levonian}}, \ and\ \bibinfo {author} {\bibfnamefont {M.~D.}\ \bibnamefont {Lukin}},\ }\bibfield  {title} {\enquote {\bibinfo {title} {Robust multi-qubit quantum network node with integrated error detection},}\ }\href {\doibase 10.1126/science.add9771} {\bibfield  {journal} {\bibinfo  {journal} {Science}\ }\textbf {\bibinfo {volume} {378}},\ \bibinfo {pages} {557--560} (\bibinfo {year} {2022})}\BibitemShut {NoStop}%
\bibitem [{\citenamefont {Nguyen}\ \emph {et~al.}(2019)\citenamefont {Nguyen}, \citenamefont {Sukachev}, \citenamefont {Bhaskar}, \citenamefont {Machielse}, \citenamefont {Levonian}, \citenamefont {Knall}, \citenamefont {Stroganov}, \citenamefont {Chia}, \citenamefont {Burek}, \citenamefont {Riedinger}, \citenamefont {Park}, \citenamefont {Lončar},\ and\ \citenamefont {Lukin}}]{nguyen_integrated_2019}%
  \BibitemOpen
  \bibfield  {author} {\bibinfo {author} {\bibfnamefont {C.~T.}\ \bibnamefont {Nguyen}}, \bibinfo {author} {\bibfnamefont {D.~D.}\ \bibnamefont {Sukachev}}, \bibinfo {author} {\bibfnamefont {M.~K.}\ \bibnamefont {Bhaskar}}, \bibinfo {author} {\bibfnamefont {B.}~\bibnamefont {Machielse}}, \bibinfo {author} {\bibfnamefont {D.~S.}\ \bibnamefont {Levonian}}, \bibinfo {author} {\bibfnamefont {E.~N.}\ \bibnamefont {Knall}}, \bibinfo {author} {\bibfnamefont {P.}~\bibnamefont {Stroganov}}, \bibinfo {author} {\bibfnamefont {C.}~\bibnamefont {Chia}}, \bibinfo {author} {\bibfnamefont {M.~J.}\ \bibnamefont {Burek}}, \bibinfo {author} {\bibfnamefont {R.}~\bibnamefont {Riedinger}}, \bibinfo {author} {\bibfnamefont {H.}~\bibnamefont {Park}}, \bibinfo {author} {\bibfnamefont {M.}~\bibnamefont {Lončar}}, \ and\ \bibinfo {author} {\bibfnamefont {M.~D.}\ \bibnamefont {Lukin}},\ }\bibfield  {title} {{\selectlanguage {en}\enquote {\bibinfo {title} {An integrated nanophotonic quantum register based on silicon-vacancy spins in
  diamond},}\ }}\href {\doibase 10.1103/PhysRevB.100.165428} {\bibfield  {journal} {\bibinfo  {journal} {Phys. Rev. B}\ }\textbf {\bibinfo {volume} {100}},\ \bibinfo {pages} {165428} (\bibinfo {year} {2019})}\BibitemShut {NoStop}%
\bibitem [{\citenamefont {Pompili}\ \emph {et~al.}(2021)\citenamefont {Pompili}, \citenamefont {Hermans}, \citenamefont {Baier}, \citenamefont {Beukers}, \citenamefont {Humphreys}, \citenamefont {Schouten}, \citenamefont {Vermeulen}, \citenamefont {Tiggelman}, \citenamefont {Martins}, \citenamefont {Dirkse}, \citenamefont {Wehner},\ and\ \citenamefont {Hanson}}]{pompili_realization_2021}%
  \BibitemOpen
  \bibfield  {author} {\bibinfo {author} {\bibfnamefont {M.}~\bibnamefont {Pompili}}, \bibinfo {author} {\bibfnamefont {S.~L.~N.}\ \bibnamefont {Hermans}}, \bibinfo {author} {\bibfnamefont {S.}~\bibnamefont {Baier}}, \bibinfo {author} {\bibfnamefont {H.~K.~C.}\ \bibnamefont {Beukers}}, \bibinfo {author} {\bibfnamefont {P.~C.}\ \bibnamefont {Humphreys}}, \bibinfo {author} {\bibfnamefont {R.~N.}\ \bibnamefont {Schouten}}, \bibinfo {author} {\bibfnamefont {R.~F.~L.}\ \bibnamefont {Vermeulen}}, \bibinfo {author} {\bibfnamefont {M.~J.}\ \bibnamefont {Tiggelman}}, \bibinfo {author} {\bibfnamefont {L.~d.~S.}\ \bibnamefont {Martins}}, \bibinfo {author} {\bibfnamefont {B.}~\bibnamefont {Dirkse}}, \bibinfo {author} {\bibfnamefont {S.}~\bibnamefont {Wehner}}, \ and\ \bibinfo {author} {\bibfnamefont {R.}~\bibnamefont {Hanson}},\ }\bibfield  {title} {\enquote {\bibinfo {title} {Realization of a multinode quantum network of remote solid-state qubits},}\ }\href {\doibase 10.1126/science.abg1919} {\bibfield  {journal}
  {\bibinfo  {journal} {Science}\ }\textbf {\bibinfo {volume} {372}},\ \bibinfo {pages} {259--264} (\bibinfo {year} {2021})}\BibitemShut {NoStop}%
\bibitem [{\citenamefont {Stolk}\ \emph {et~al.}(2024)\citenamefont {Stolk}, \citenamefont {Enden}, \citenamefont {Slater}, \citenamefont {Raa-Derckx}, \citenamefont {Botma}, \citenamefont {Rantwijk}, \citenamefont {Biemond}, \citenamefont {Hagen}, \citenamefont {Herfst}, \citenamefont {Koek}, \citenamefont {Meskers}, \citenamefont {Vollmer}, \citenamefont {Zwet}, \citenamefont {Markham}, \citenamefont {Edmonds}, \citenamefont {Geus}, \citenamefont {Elsen}, \citenamefont {Jungbluth}, \citenamefont {Haefner}, \citenamefont {Tresp}, \citenamefont {Stuhler}, \citenamefont {Ritter},\ and\ \citenamefont {Hanson}}]{stolk_metropolitan-scale_2024}%
  \BibitemOpen
  \bibfield  {author} {\bibinfo {author} {\bibfnamefont {A.~J.}\ \bibnamefont {Stolk}}, \bibinfo {author} {\bibfnamefont {K.~L. v.~d.}\ \bibnamefont {Enden}}, \bibinfo {author} {\bibfnamefont {M.-C.}\ \bibnamefont {Slater}}, \bibinfo {author} {\bibfnamefont {I.~t.}\ \bibnamefont {Raa-Derckx}}, \bibinfo {author} {\bibfnamefont {P.}~\bibnamefont {Botma}}, \bibinfo {author} {\bibfnamefont {J.~v.}\ \bibnamefont {Rantwijk}}, \bibinfo {author} {\bibfnamefont {J.~J.~B.}\ \bibnamefont {Biemond}}, \bibinfo {author} {\bibfnamefont {R.~A.~J.}\ \bibnamefont {Hagen}}, \bibinfo {author} {\bibfnamefont {R.~W.}\ \bibnamefont {Herfst}}, \bibinfo {author} {\bibfnamefont {W.~D.}\ \bibnamefont {Koek}}, \bibinfo {author} {\bibfnamefont {A.~J.~H.}\ \bibnamefont {Meskers}}, \bibinfo {author} {\bibfnamefont {R.}~\bibnamefont {Vollmer}}, \bibinfo {author} {\bibfnamefont {E.~J.~v.}\ \bibnamefont {Zwet}}, \bibinfo {author} {\bibfnamefont {M.}~\bibnamefont {Markham}}, \bibinfo {author} {\bibfnamefont {A.~M.}\ \bibnamefont {Edmonds}},
  \bibinfo {author} {\bibfnamefont {J.~F.}\ \bibnamefont {Geus}}, \bibinfo {author} {\bibfnamefont {F.}~\bibnamefont {Elsen}}, \bibinfo {author} {\bibfnamefont {B.}~\bibnamefont {Jungbluth}}, \bibinfo {author} {\bibfnamefont {C.}~\bibnamefont {Haefner}}, \bibinfo {author} {\bibfnamefont {C.}~\bibnamefont {Tresp}}, \bibinfo {author} {\bibfnamefont {J.}~\bibnamefont {Stuhler}}, \bibinfo {author} {\bibfnamefont {S.}~\bibnamefont {Ritter}}, \ and\ \bibinfo {author} {\bibfnamefont {R.}~\bibnamefont {Hanson}},\ }\bibfield  {title} {\enquote {\bibinfo {title} {Metropolitan-scale heralded entanglement of solid-state qubits},}\ }\href {\doibase 10.1126/sciadv.adp6442} {\bibfield  {journal} {\bibinfo  {journal} {Science Advances}\ } (\bibinfo {year} {2024}),\ 10.1126/sciadv.adp6442}\BibitemShut {NoStop}%
\bibitem [{\citenamefont {Li}\ \emph {et~al.}(2024{\natexlab{a}})\citenamefont {Li}, \citenamefont {Santis}, \citenamefont {Harris}, \citenamefont {Chen}, \citenamefont {Gao}, \citenamefont {Christen}, \citenamefont {Choi}, \citenamefont {Trusheim}, \citenamefont {Song}, \citenamefont {Errando-Herranz}, \citenamefont {Du}, \citenamefont {Hu}, \citenamefont {Clark}, \citenamefont {Ibrahim}, \citenamefont {Gilbert}, \citenamefont {Han},\ and\ \citenamefont {Englund}}]{li_heterogeneous_2024}%
  \BibitemOpen
  \bibfield  {author} {\bibinfo {author} {\bibfnamefont {L.}~\bibnamefont {Li}}, \bibinfo {author} {\bibfnamefont {L.~D.}\ \bibnamefont {Santis}}, \bibinfo {author} {\bibfnamefont {I.~B.~W.}\ \bibnamefont {Harris}}, \bibinfo {author} {\bibfnamefont {K.~C.}\ \bibnamefont {Chen}}, \bibinfo {author} {\bibfnamefont {Y.}~\bibnamefont {Gao}}, \bibinfo {author} {\bibfnamefont {I.}~\bibnamefont {Christen}}, \bibinfo {author} {\bibfnamefont {H.}~\bibnamefont {Choi}}, \bibinfo {author} {\bibfnamefont {M.}~\bibnamefont {Trusheim}}, \bibinfo {author} {\bibfnamefont {Y.}~\bibnamefont {Song}}, \bibinfo {author} {\bibfnamefont {C.}~\bibnamefont {Errando-Herranz}}, \bibinfo {author} {\bibfnamefont {J.}~\bibnamefont {Du}}, \bibinfo {author} {\bibfnamefont {Y.}~\bibnamefont {Hu}}, \bibinfo {author} {\bibfnamefont {G.}~\bibnamefont {Clark}}, \bibinfo {author} {\bibfnamefont {M.~I.}\ \bibnamefont {Ibrahim}}, \bibinfo {author} {\bibfnamefont {G.}~\bibnamefont {Gilbert}}, \bibinfo {author} {\bibfnamefont {R.}~\bibnamefont {Han}},
  \ and\ \bibinfo {author} {\bibfnamefont {D.}~\bibnamefont {Englund}},\ }\bibfield  {title} {\enquote {\bibinfo {title} {Heterogeneous integration of spin–photon interfaces with a {CMOS} platform},}\ }\href {\doibase 10.1038/s41586-024-07371-7} {\bibfield  {journal} {\bibinfo  {journal} {Nature}\ }\textbf {\bibinfo {volume} {630}},\ \bibinfo {pages} {70--76} (\bibinfo {year} {2024}{\natexlab{a}})}\BibitemShut {NoStop}%
\bibitem [{\citenamefont {Bhaskar}\ \emph {et~al.}(2020)\citenamefont {Bhaskar}, \citenamefont {Riedinger}, \citenamefont {Machielse}, \citenamefont {Levonian}, \citenamefont {Nguyen}, \citenamefont {Knall}, \citenamefont {Park}, \citenamefont {Englund}, \citenamefont {Lončar}, \citenamefont {Sukachev},\ and\ \citenamefont {Lukin}}]{bhaskar_experimental_2020}%
  \BibitemOpen
  \bibfield  {author} {\bibinfo {author} {\bibfnamefont {M.~K.}\ \bibnamefont {Bhaskar}}, \bibinfo {author} {\bibfnamefont {R.}~\bibnamefont {Riedinger}}, \bibinfo {author} {\bibfnamefont {B.}~\bibnamefont {Machielse}}, \bibinfo {author} {\bibfnamefont {D.~S.}\ \bibnamefont {Levonian}}, \bibinfo {author} {\bibfnamefont {C.~T.}\ \bibnamefont {Nguyen}}, \bibinfo {author} {\bibfnamefont {E.~N.}\ \bibnamefont {Knall}}, \bibinfo {author} {\bibfnamefont {H.}~\bibnamefont {Park}}, \bibinfo {author} {\bibfnamefont {D.}~\bibnamefont {Englund}}, \bibinfo {author} {\bibfnamefont {M.}~\bibnamefont {Lončar}}, \bibinfo {author} {\bibfnamefont {D.~D.}\ \bibnamefont {Sukachev}}, \ and\ \bibinfo {author} {\bibfnamefont {M.~D.}\ \bibnamefont {Lukin}},\ }\bibfield  {title} {\enquote {\bibinfo {title} {Experimental demonstration of memory-enhanced quantum communication},}\ }\href {\doibase 10.1038/s41586-020-2103-5} {\bibfield  {journal} {\bibinfo  {journal} {Nature}\ }\textbf {\bibinfo {volume} {580}},\ \bibinfo {pages}
  {60--64} (\bibinfo {year} {2020})}\BibitemShut {NoStop}%
\bibitem [{\citenamefont {Hensen}\ \emph {et~al.}(2015)\citenamefont {Hensen}, \citenamefont {Bernien}, \citenamefont {Dréau}, \citenamefont {Reiserer}, \citenamefont {Kalb}, \citenamefont {Blok}, \citenamefont {Ruitenberg}, \citenamefont {Vermeulen}, \citenamefont {Schouten}, \citenamefont {Abellán}, \citenamefont {Amaya}, \citenamefont {Pruneri}, \citenamefont {Mitchell}, \citenamefont {Markham}, \citenamefont {Twitchen}, \citenamefont {Elkouss}, \citenamefont {Wehner}, \citenamefont {Taminiau},\ and\ \citenamefont {Hanson}}]{hensen_loophole-free_2015}%
  \BibitemOpen
  \bibfield  {author} {\bibinfo {author} {\bibfnamefont {B.}~\bibnamefont {Hensen}}, \bibinfo {author} {\bibfnamefont {H.}~\bibnamefont {Bernien}}, \bibinfo {author} {\bibfnamefont {A.~E.}\ \bibnamefont {Dréau}}, \bibinfo {author} {\bibfnamefont {A.}~\bibnamefont {Reiserer}}, \bibinfo {author} {\bibfnamefont {N.}~\bibnamefont {Kalb}}, \bibinfo {author} {\bibfnamefont {M.~S.}\ \bibnamefont {Blok}}, \bibinfo {author} {\bibfnamefont {J.}~\bibnamefont {Ruitenberg}}, \bibinfo {author} {\bibfnamefont {R.~F.~L.}\ \bibnamefont {Vermeulen}}, \bibinfo {author} {\bibfnamefont {R.~N.}\ \bibnamefont {Schouten}}, \bibinfo {author} {\bibfnamefont {C.}~\bibnamefont {Abellán}}, \bibinfo {author} {\bibfnamefont {W.}~\bibnamefont {Amaya}}, \bibinfo {author} {\bibfnamefont {V.}~\bibnamefont {Pruneri}}, \bibinfo {author} {\bibfnamefont {M.~W.}\ \bibnamefont {Mitchell}}, \bibinfo {author} {\bibfnamefont {M.}~\bibnamefont {Markham}}, \bibinfo {author} {\bibfnamefont {D.~J.}\ \bibnamefont {Twitchen}}, \bibinfo {author}
  {\bibfnamefont {D.}~\bibnamefont {Elkouss}}, \bibinfo {author} {\bibfnamefont {S.}~\bibnamefont {Wehner}}, \bibinfo {author} {\bibfnamefont {T.~H.}\ \bibnamefont {Taminiau}}, \ and\ \bibinfo {author} {\bibfnamefont {R.}~\bibnamefont {Hanson}},\ }\bibfield  {title} {\enquote {\bibinfo {title} {Loophole-free {Bell} inequality violation using electron spins separated by 1.3 kilometres},}\ }\href {\doibase 10.1038/nature15759} {\bibfield  {journal} {\bibinfo  {journal} {Nature}\ }\textbf {\bibinfo {volume} {526}},\ \bibinfo {pages} {682--686} (\bibinfo {year} {2015})}\BibitemShut {NoStop}%
\bibitem [{\citenamefont {Wan}\ \emph {et~al.}(2020)\citenamefont {Wan}, \citenamefont {Lu}, \citenamefont {Chen}, \citenamefont {Walsh}, \citenamefont {Trusheim}, \citenamefont {De~Santis}, \citenamefont {Bersin}, \citenamefont {Harris}, \citenamefont {Mouradian}, \citenamefont {Christen}, \citenamefont {Bielejec},\ and\ \citenamefont {Englund}}]{wan_large-scale_2020}%
  \BibitemOpen
  \bibfield  {author} {\bibinfo {author} {\bibfnamefont {N.~H.}\ \bibnamefont {Wan}}, \bibinfo {author} {\bibfnamefont {T.-J.}\ \bibnamefont {Lu}}, \bibinfo {author} {\bibfnamefont {K.~C.}\ \bibnamefont {Chen}}, \bibinfo {author} {\bibfnamefont {M.~P.}\ \bibnamefont {Walsh}}, \bibinfo {author} {\bibfnamefont {M.~E.}\ \bibnamefont {Trusheim}}, \bibinfo {author} {\bibfnamefont {L.}~\bibnamefont {De~Santis}}, \bibinfo {author} {\bibfnamefont {E.~A.}\ \bibnamefont {Bersin}}, \bibinfo {author} {\bibfnamefont {I.~B.}\ \bibnamefont {Harris}}, \bibinfo {author} {\bibfnamefont {S.~L.}\ \bibnamefont {Mouradian}}, \bibinfo {author} {\bibfnamefont {I.~R.}\ \bibnamefont {Christen}}, \bibinfo {author} {\bibfnamefont {E.~S.}\ \bibnamefont {Bielejec}}, \ and\ \bibinfo {author} {\bibfnamefont {D.}~\bibnamefont {Englund}},\ }\bibfield  {title} {\enquote {\bibinfo {title} {Large-scale integration of artificial atoms in hybrid photonic circuits},}\ }\href {\doibase 10.1038/s41586-020-2441-3} {\bibfield  {journal} {\bibinfo
  {journal} {Nature}\ }\textbf {\bibinfo {volume} {583}},\ \bibinfo {pages} {226--231} (\bibinfo {year} {2020})}\BibitemShut {NoStop}%
\bibitem [{\citenamefont {Heiler}\ \emph {et~al.}(2024)\citenamefont {Heiler}, \citenamefont {Körber}, \citenamefont {Hesselmeier}, \citenamefont {Kuna}, \citenamefont {Stöhr}, \citenamefont {Fuchs}, \citenamefont {Ghezellou}, \citenamefont {Ul-Hassan}, \citenamefont {Knolle}, \citenamefont {Becher}, \citenamefont {Kaiser},\ and\ \citenamefont {Wrachtrup}}]{heiler_spectral_2024}%
  \BibitemOpen
  \bibfield  {author} {\bibinfo {author} {\bibfnamefont {J.}~\bibnamefont {Heiler}}, \bibinfo {author} {\bibfnamefont {J.}~\bibnamefont {Körber}}, \bibinfo {author} {\bibfnamefont {E.}~\bibnamefont {Hesselmeier}}, \bibinfo {author} {\bibfnamefont {P.}~\bibnamefont {Kuna}}, \bibinfo {author} {\bibfnamefont {R.}~\bibnamefont {Stöhr}}, \bibinfo {author} {\bibfnamefont {P.}~\bibnamefont {Fuchs}}, \bibinfo {author} {\bibfnamefont {M.}~\bibnamefont {Ghezellou}}, \bibinfo {author} {\bibfnamefont {J.}~\bibnamefont {Ul-Hassan}}, \bibinfo {author} {\bibfnamefont {W.}~\bibnamefont {Knolle}}, \bibinfo {author} {\bibfnamefont {C.}~\bibnamefont {Becher}}, \bibinfo {author} {\bibfnamefont {F.}~\bibnamefont {Kaiser}}, \ and\ \bibinfo {author} {\bibfnamefont {J.}~\bibnamefont {Wrachtrup}},\ }\bibfield  {title} {\enquote {\bibinfo {title} {Spectral stability of {V2} centres in sub-micron {4H}-{SiC} membranes},}\ }\href {\doibase 10.1038/s41535-024-00644-4} {\bibfield  {journal} {\bibinfo  {journal} {npj Quantum Mater.}\
  }\textbf {\bibinfo {volume} {9}},\ \bibinfo {pages} {34} (\bibinfo {year} {2024})}\BibitemShut {NoStop}%
\bibitem [{\citenamefont {Simmons}(2024)}]{simmons_scalable_2024}%
  \BibitemOpen
  \bibfield  {author} {\bibinfo {author} {\bibfnamefont {S.}~\bibnamefont {Simmons}},\ }\bibfield  {title} {\enquote {\bibinfo {title} {Scalable {Fault}-{Tolerant} {Quantum} {Technologies} with {Silicon} {Color} {Centers}},}\ }\href {\doibase 10.1103/PRXQuantum.5.010102} {\bibfield  {journal} {\bibinfo  {journal} {PRX Quantum}\ }\textbf {\bibinfo {volume} {5}},\ \bibinfo {pages} {010102} (\bibinfo {year} {2024})}\BibitemShut {NoStop}%
\bibitem [{\citenamefont {Bhaskar}\ \emph {et~al.}(2017)\citenamefont {Bhaskar}, \citenamefont {Sukachev}, \citenamefont {Sipahigil}, \citenamefont {Evans}, \citenamefont {Burek}, \citenamefont {Nguyen}, \citenamefont {Rogers}, \citenamefont {Siyushev}, \citenamefont {Metsch}, \citenamefont {Park}, \citenamefont {Jelezko}, \citenamefont {Lončar},\ and\ \citenamefont {Lukin}}]{bhaskar_quantum_2017}%
  \BibitemOpen
  \bibfield  {author} {\bibinfo {author} {\bibfnamefont {M.}~\bibnamefont {Bhaskar}}, \bibinfo {author} {\bibfnamefont {D.}~\bibnamefont {Sukachev}}, \bibinfo {author} {\bibfnamefont {A.}~\bibnamefont {Sipahigil}}, \bibinfo {author} {\bibfnamefont {R.}~\bibnamefont {Evans}}, \bibinfo {author} {\bibfnamefont {M.}~\bibnamefont {Burek}}, \bibinfo {author} {\bibfnamefont {C.}~\bibnamefont {Nguyen}}, \bibinfo {author} {\bibfnamefont {L.}~\bibnamefont {Rogers}}, \bibinfo {author} {\bibfnamefont {P.}~\bibnamefont {Siyushev}}, \bibinfo {author} {\bibfnamefont {M.}~\bibnamefont {Metsch}}, \bibinfo {author} {\bibfnamefont {H.}~\bibnamefont {Park}}, \bibinfo {author} {\bibfnamefont {F.}~\bibnamefont {Jelezko}}, \bibinfo {author} {\bibfnamefont {M.}~\bibnamefont {Lončar}}, \ and\ \bibinfo {author} {\bibfnamefont {M.}~\bibnamefont {Lukin}},\ }\bibfield  {title} {\enquote {\bibinfo {title} {Quantum {Nonlinear} {Optics} with a {Germanium}-{Vacancy} {Color} {Center} in a {Nanoscale} {Diamond} {Waveguide}},}\ }\href
  {\doibase 10.1103/PhysRevLett.118.223603} {\bibfield  {journal} {\bibinfo  {journal} {Phys. Rev. Lett.}\ }\textbf {\bibinfo {volume} {118}},\ \bibinfo {pages} {223603} (\bibinfo {year} {2017})}\BibitemShut {NoStop}%
\bibitem [{\citenamefont {Trusheim}\ \emph {et~al.}(2020)\citenamefont {Trusheim}, \citenamefont {Pingault}, \citenamefont {Wan}, \citenamefont {Gündoğan}, \citenamefont {De~Santis}, \citenamefont {Debroux}, \citenamefont {Gangloff}, \citenamefont {Purser}, \citenamefont {Chen}, \citenamefont {Walsh}, \citenamefont {Rose}, \citenamefont {Becker}, \citenamefont {Lienhard}, \citenamefont {Bersin}, \citenamefont {Paradeisanos}, \citenamefont {Wang}, \citenamefont {Lyzwa}, \citenamefont {Montblanch}, \citenamefont {Malladi}, \citenamefont {Bakhru}, \citenamefont {Ferrari}, \citenamefont {Walmsley}, \citenamefont {Atatüre},\ and\ \citenamefont {Englund}}]{trusheim_transform-limited_2020}%
  \BibitemOpen
  \bibfield  {author} {\bibinfo {author} {\bibfnamefont {M.~E.}\ \bibnamefont {Trusheim}}, \bibinfo {author} {\bibfnamefont {B.}~\bibnamefont {Pingault}}, \bibinfo {author} {\bibfnamefont {N.~H.}\ \bibnamefont {Wan}}, \bibinfo {author} {\bibfnamefont {M.}~\bibnamefont {Gündoğan}}, \bibinfo {author} {\bibfnamefont {L.}~\bibnamefont {De~Santis}}, \bibinfo {author} {\bibfnamefont {R.}~\bibnamefont {Debroux}}, \bibinfo {author} {\bibfnamefont {D.}~\bibnamefont {Gangloff}}, \bibinfo {author} {\bibfnamefont {C.}~\bibnamefont {Purser}}, \bibinfo {author} {\bibfnamefont {K.~C.}\ \bibnamefont {Chen}}, \bibinfo {author} {\bibfnamefont {M.}~\bibnamefont {Walsh}}, \bibinfo {author} {\bibfnamefont {J.~J.}\ \bibnamefont {Rose}}, \bibinfo {author} {\bibfnamefont {J.~N.}\ \bibnamefont {Becker}}, \bibinfo {author} {\bibfnamefont {B.}~\bibnamefont {Lienhard}}, \bibinfo {author} {\bibfnamefont {E.}~\bibnamefont {Bersin}}, \bibinfo {author} {\bibfnamefont {I.}~\bibnamefont {Paradeisanos}}, \bibinfo {author} {\bibfnamefont
  {G.}~\bibnamefont {Wang}}, \bibinfo {author} {\bibfnamefont {D.}~\bibnamefont {Lyzwa}}, \bibinfo {author} {\bibfnamefont {A.~R.-P.}\ \bibnamefont {Montblanch}}, \bibinfo {author} {\bibfnamefont {G.}~\bibnamefont {Malladi}}, \bibinfo {author} {\bibfnamefont {H.}~\bibnamefont {Bakhru}}, \bibinfo {author} {\bibfnamefont {A.~C.}\ \bibnamefont {Ferrari}}, \bibinfo {author} {\bibfnamefont {I.~A.}\ \bibnamefont {Walmsley}}, \bibinfo {author} {\bibfnamefont {M.}~\bibnamefont {Atatüre}}, \ and\ \bibinfo {author} {\bibfnamefont {D.}~\bibnamefont {Englund}},\ }\bibfield  {title} {\enquote {\bibinfo {title} {Transform-{Limited} {Photons} {From} a {Coherent} {Tin}-{Vacancy} {Spin} in {Diamond}},}\ }\href {\doibase 10.1103/PhysRevLett.124.023602} {\bibfield  {journal} {\bibinfo  {journal} {Phys. Rev. Lett.}\ }\textbf {\bibinfo {volume} {124}},\ \bibinfo {pages} {023602} (\bibinfo {year} {2020})}\BibitemShut {NoStop}%
\bibitem [{\citenamefont {Brevoord}\ \emph {et~al.}(2024{\natexlab{a}})\citenamefont {Brevoord}, \citenamefont {De~Santis}, \citenamefont {Yamamoto}, \citenamefont {Pasini}, \citenamefont {Codreanu}, \citenamefont {Turan}, \citenamefont {Beukers}, \citenamefont {Waas},\ and\ \citenamefont {Hanson}}]{brevoord_heralded_2024}%
  \BibitemOpen
  \bibfield  {author} {\bibinfo {author} {\bibfnamefont {J.~M.}\ \bibnamefont {Brevoord}}, \bibinfo {author} {\bibfnamefont {L.}~\bibnamefont {De~Santis}}, \bibinfo {author} {\bibfnamefont {T.}~\bibnamefont {Yamamoto}}, \bibinfo {author} {\bibfnamefont {M.}~\bibnamefont {Pasini}}, \bibinfo {author} {\bibfnamefont {N.}~\bibnamefont {Codreanu}}, \bibinfo {author} {\bibfnamefont {T.}~\bibnamefont {Turan}}, \bibinfo {author} {\bibfnamefont {H.~K.}\ \bibnamefont {Beukers}}, \bibinfo {author} {\bibfnamefont {C.}~\bibnamefont {Waas}}, \ and\ \bibinfo {author} {\bibfnamefont {R.}~\bibnamefont {Hanson}},\ }\bibfield  {title} {\enquote {\bibinfo {title} {Heralded initialization of charge state and optical-transition frequency of diamond tin-vacancy centers},}\ }\href {\doibase 10.1103/PhysRevApplied.21.054047} {\bibfield  {journal} {\bibinfo  {journal} {Phys. Rev. Applied}\ }\textbf {\bibinfo {volume} {21}},\ \bibinfo {pages} {054047} (\bibinfo {year} {2024}{\natexlab{a}})}\BibitemShut {NoStop}%
\bibitem [{\citenamefont {Rugar}\ \emph {et~al.}(2021)\citenamefont {Rugar}, \citenamefont {Aghaeimeibodi}, \citenamefont {Riedel}, \citenamefont {Dory}, \citenamefont {Lu}, \citenamefont {McQuade}, \citenamefont {Shen}, \citenamefont {Melosh},\ and\ \citenamefont {Vučković}}]{rugar_quantum_2021}%
  \BibitemOpen
  \bibfield  {author} {\bibinfo {author} {\bibfnamefont {A.~E.}\ \bibnamefont {Rugar}}, \bibinfo {author} {\bibfnamefont {S.}~\bibnamefont {Aghaeimeibodi}}, \bibinfo {author} {\bibfnamefont {D.}~\bibnamefont {Riedel}}, \bibinfo {author} {\bibfnamefont {C.}~\bibnamefont {Dory}}, \bibinfo {author} {\bibfnamefont {H.}~\bibnamefont {Lu}}, \bibinfo {author} {\bibfnamefont {P.~J.}\ \bibnamefont {McQuade}}, \bibinfo {author} {\bibfnamefont {Z.-X.}\ \bibnamefont {Shen}}, \bibinfo {author} {\bibfnamefont {N.~A.}\ \bibnamefont {Melosh}}, \ and\ \bibinfo {author} {\bibfnamefont {J.}~\bibnamefont {Vučković}},\ }\bibfield  {title} {\enquote {\bibinfo {title} {Quantum {Photonic} {Interface} for {Tin}-{Vacancy} {Centers} in {Diamond}},}\ }\href {\doibase 10.1103/PhysRevX.11.031021} {\bibfield  {journal} {\bibinfo  {journal} {Phys. Rev. X}\ }\textbf {\bibinfo {volume} {11}},\ \bibinfo {pages} {031021} (\bibinfo {year} {2021})}\BibitemShut {NoStop}%
\bibitem [{\citenamefont {Iwasaki}\ \emph {et~al.}(2017)\citenamefont {Iwasaki}, \citenamefont {Miyamoto}, \citenamefont {Taniguchi}, \citenamefont {Siyushev}, \citenamefont {Metsch}, \citenamefont {Jelezko},\ and\ \citenamefont {Hatano}}]{iwasaki_tin-vacancy_2017}%
  \BibitemOpen
  \bibfield  {author} {\bibinfo {author} {\bibfnamefont {T.}~\bibnamefont {Iwasaki}}, \bibinfo {author} {\bibfnamefont {Y.}~\bibnamefont {Miyamoto}}, \bibinfo {author} {\bibfnamefont {T.}~\bibnamefont {Taniguchi}}, \bibinfo {author} {\bibfnamefont {P.}~\bibnamefont {Siyushev}}, \bibinfo {author} {\bibfnamefont {M.~H.}\ \bibnamefont {Metsch}}, \bibinfo {author} {\bibfnamefont {F.}~\bibnamefont {Jelezko}}, \ and\ \bibinfo {author} {\bibfnamefont {M.}~\bibnamefont {Hatano}},\ }\bibfield  {title} {\enquote {\bibinfo {title} {Tin-{Vacancy} {Quantum} {Emitters} in {Diamond}},}\ }\href {\doibase 10.1103/PhysRevLett.119.253601} {\bibfield  {journal} {\bibinfo  {journal} {Phys. Rev. Lett.}\ }\textbf {\bibinfo {volume} {119}},\ \bibinfo {pages} {253601} (\bibinfo {year} {2017})}\BibitemShut {NoStop}%
\bibitem [{\citenamefont {Görlitz}\ \emph {et~al.}(2020)\citenamefont {Görlitz}, \citenamefont {Herrmann}, \citenamefont {Thiering}, \citenamefont {Fuchs}, \citenamefont {Gandil}, \citenamefont {Iwasaki}, \citenamefont {Taniguchi}, \citenamefont {Kieschnick}, \citenamefont {Meijer}, \citenamefont {Hatano}, \citenamefont {Gali},\ and\ \citenamefont {Becher}}]{gorlitz_spectroscopic_2020}%
  \BibitemOpen
  \bibfield  {author} {\bibinfo {author} {\bibfnamefont {J.}~\bibnamefont {Görlitz}}, \bibinfo {author} {\bibfnamefont {D.}~\bibnamefont {Herrmann}}, \bibinfo {author} {\bibfnamefont {G.}~\bibnamefont {Thiering}}, \bibinfo {author} {\bibfnamefont {P.}~\bibnamefont {Fuchs}}, \bibinfo {author} {\bibfnamefont {M.}~\bibnamefont {Gandil}}, \bibinfo {author} {\bibfnamefont {T.}~\bibnamefont {Iwasaki}}, \bibinfo {author} {\bibfnamefont {T.}~\bibnamefont {Taniguchi}}, \bibinfo {author} {\bibfnamefont {M.}~\bibnamefont {Kieschnick}}, \bibinfo {author} {\bibfnamefont {J.}~\bibnamefont {Meijer}}, \bibinfo {author} {\bibfnamefont {M.}~\bibnamefont {Hatano}}, \bibinfo {author} {\bibfnamefont {A.}~\bibnamefont {Gali}}, \ and\ \bibinfo {author} {\bibfnamefont {C.}~\bibnamefont {Becher}},\ }\bibfield  {title} {\enquote {\bibinfo {title} {Spectroscopic investigations of negatively charged tin-vacancy centres in diamond},}\ }\href {\doibase 10.1088/1367-2630/ab6631} {\bibfield  {journal} {\bibinfo  {journal} {New J. Phys.}\
  }\textbf {\bibinfo {volume} {22}},\ \bibinfo {pages} {013048} (\bibinfo {year} {2020})}\BibitemShut {NoStop}%
\bibitem [{\citenamefont {Rosenthal}\ \emph {et~al.}(2023)\citenamefont {Rosenthal}, \citenamefont {Anderson}, \citenamefont {Kleidermacher}, \citenamefont {Stein}, \citenamefont {Lee}, \citenamefont {Grzesik}, \citenamefont {Scuri}, \citenamefont {Rugar}, \citenamefont {Riedel}, \citenamefont {Aghaeimeibodi}, \citenamefont {Ahn}, \citenamefont {Van~Gasse},\ and\ \citenamefont {Vučković}}]{rosenthal_microwave_2023}%
  \BibitemOpen
  \bibfield  {author} {\bibinfo {author} {\bibfnamefont {E.~I.}\ \bibnamefont {Rosenthal}}, \bibinfo {author} {\bibfnamefont {C.~P.}\ \bibnamefont {Anderson}}, \bibinfo {author} {\bibfnamefont {H.~C.}\ \bibnamefont {Kleidermacher}}, \bibinfo {author} {\bibfnamefont {A.~J.}\ \bibnamefont {Stein}}, \bibinfo {author} {\bibfnamefont {H.}~\bibnamefont {Lee}}, \bibinfo {author} {\bibfnamefont {J.}~\bibnamefont {Grzesik}}, \bibinfo {author} {\bibfnamefont {G.}~\bibnamefont {Scuri}}, \bibinfo {author} {\bibfnamefont {A.~E.}\ \bibnamefont {Rugar}}, \bibinfo {author} {\bibfnamefont {D.}~\bibnamefont {Riedel}}, \bibinfo {author} {\bibfnamefont {S.}~\bibnamefont {Aghaeimeibodi}}, \bibinfo {author} {\bibfnamefont {G.~H.}\ \bibnamefont {Ahn}}, \bibinfo {author} {\bibfnamefont {K.}~\bibnamefont {Van~Gasse}}, \ and\ \bibinfo {author} {\bibfnamefont {J.}~\bibnamefont {Vučković}},\ }\bibfield  {title} {\enquote {\bibinfo {title} {Microwave {Spin} {Control} of a {Tin}-{Vacancy} {Qubit} in {Diamond}},}\ }\href {\doibase
  10.1103/PhysRevX.13.031022} {\bibfield  {journal} {\bibinfo  {journal} {Phys. Rev. X}\ }\textbf {\bibinfo {volume} {13}},\ \bibinfo {pages} {031022} (\bibinfo {year} {2023})}\BibitemShut {NoStop}%
\bibitem [{\citenamefont {Karapatzakis}\ \emph {et~al.}(2024)\citenamefont {Karapatzakis}, \citenamefont {Resch}, \citenamefont {Schrodin}, \citenamefont {Fuchs}, \citenamefont {Kieschnick}, \citenamefont {Heupel}, \citenamefont {Kussi}, \citenamefont {Sürgers}, \citenamefont {Popov}, \citenamefont {Meijer}, \citenamefont {Becher}, \citenamefont {Wernsdorfer},\ and\ \citenamefont {Hunger}}]{karapatzakis_microwave_2024}%
  \BibitemOpen
  \bibfield  {author} {\bibinfo {author} {\bibfnamefont {I.}~\bibnamefont {Karapatzakis}}, \bibinfo {author} {\bibfnamefont {J.}~\bibnamefont {Resch}}, \bibinfo {author} {\bibfnamefont {M.}~\bibnamefont {Schrodin}}, \bibinfo {author} {\bibfnamefont {P.}~\bibnamefont {Fuchs}}, \bibinfo {author} {\bibfnamefont {M.}~\bibnamefont {Kieschnick}}, \bibinfo {author} {\bibfnamefont {J.}~\bibnamefont {Heupel}}, \bibinfo {author} {\bibfnamefont {L.}~\bibnamefont {Kussi}}, \bibinfo {author} {\bibfnamefont {C.}~\bibnamefont {Sürgers}}, \bibinfo {author} {\bibfnamefont {C.}~\bibnamefont {Popov}}, \bibinfo {author} {\bibfnamefont {J.}~\bibnamefont {Meijer}}, \bibinfo {author} {\bibfnamefont {C.}~\bibnamefont {Becher}}, \bibinfo {author} {\bibfnamefont {W.}~\bibnamefont {Wernsdorfer}}, \ and\ \bibinfo {author} {\bibfnamefont {D.}~\bibnamefont {Hunger}},\ }\bibfield  {title} {\enquote {\bibinfo {title} {Microwave {Control} of the {Tin}-{Vacancy} {Spin} {Qubit} in {Diamond} with a {Superconducting} {Waveguide}},}\ }\href
  {\doibase 10.1103/PhysRevX.14.031036} {\bibfield  {journal} {\bibinfo  {journal} {Phys. Rev. X}\ }\textbf {\bibinfo {volume} {14}},\ \bibinfo {pages} {031036} (\bibinfo {year} {2024})}\BibitemShut {NoStop}%
\bibitem [{\citenamefont {Guo}\ \emph {et~al.}(2023)\citenamefont {Guo}, \citenamefont {Stramma}, \citenamefont {Li}, \citenamefont {Roth}, \citenamefont {Huang}, \citenamefont {Jin}, \citenamefont {Parker}, \citenamefont {Arjona~Martínez}, \citenamefont {Shofer}, \citenamefont {Michaels}, \citenamefont {Purser}, \citenamefont {Appel}, \citenamefont {Alexeev}, \citenamefont {Liu}, \citenamefont {Ferrari}, \citenamefont {Awschalom}, \citenamefont {Delegan}, \citenamefont {Pingault}, \citenamefont {Galli}, \citenamefont {Heremans}, \citenamefont {Atatüre},\ and\ \citenamefont {High}}]{guo_microwave-based_2023}%
  \BibitemOpen
  \bibfield  {author} {\bibinfo {author} {\bibfnamefont {X.}~\bibnamefont {Guo}}, \bibinfo {author} {\bibfnamefont {A.~M.}\ \bibnamefont {Stramma}}, \bibinfo {author} {\bibfnamefont {Z.}~\bibnamefont {Li}}, \bibinfo {author} {\bibfnamefont {W.~G.}\ \bibnamefont {Roth}}, \bibinfo {author} {\bibfnamefont {B.}~\bibnamefont {Huang}}, \bibinfo {author} {\bibfnamefont {Y.}~\bibnamefont {Jin}}, \bibinfo {author} {\bibfnamefont {R.~A.}\ \bibnamefont {Parker}}, \bibinfo {author} {\bibfnamefont {J.}~\bibnamefont {Arjona~Martínez}}, \bibinfo {author} {\bibfnamefont {N.}~\bibnamefont {Shofer}}, \bibinfo {author} {\bibfnamefont {C.~P.}\ \bibnamefont {Michaels}}, \bibinfo {author} {\bibfnamefont {C.~P.}\ \bibnamefont {Purser}}, \bibinfo {author} {\bibfnamefont {M.~H.}\ \bibnamefont {Appel}}, \bibinfo {author} {\bibfnamefont {E.~M.}\ \bibnamefont {Alexeev}}, \bibinfo {author} {\bibfnamefont {T.}~\bibnamefont {Liu}}, \bibinfo {author} {\bibfnamefont {A.~C.}\ \bibnamefont {Ferrari}}, \bibinfo {author} {\bibfnamefont {D.~D.}\
  \bibnamefont {Awschalom}}, \bibinfo {author} {\bibfnamefont {N.}~\bibnamefont {Delegan}}, \bibinfo {author} {\bibfnamefont {B.}~\bibnamefont {Pingault}}, \bibinfo {author} {\bibfnamefont {G.}~\bibnamefont {Galli}}, \bibinfo {author} {\bibfnamefont {F.~J.}\ \bibnamefont {Heremans}}, \bibinfo {author} {\bibfnamefont {M.}~\bibnamefont {Atatüre}}, \ and\ \bibinfo {author} {\bibfnamefont {A.~A.}\ \bibnamefont {High}},\ }\bibfield  {title} {\enquote {\bibinfo {title} {Microwave-{Based} {Quantum} {Control} and {Coherence} {Protection} of {Tin}-{Vacancy} {Spin} {Qubits} in a {Strain}-{Tuned} {Diamond}-{Membrane} {Heterostructure}},}\ }\href {\doibase 10.1103/PhysRevX.13.041037} {\bibfield  {journal} {\bibinfo  {journal} {Phys. Rev. X}\ }\textbf {\bibinfo {volume} {13}},\ \bibinfo {pages} {041037} (\bibinfo {year} {2023})}\BibitemShut {NoStop}%
\bibitem [{\citenamefont {Beukers}\ \emph {et~al.}(2024)\citenamefont {Beukers}, \citenamefont {Waas}, \citenamefont {Pasini}, \citenamefont {Ommen}, \citenamefont {Ademi}, \citenamefont {Iuliano}, \citenamefont {Codreanu}, \citenamefont {Brevoord}, \citenamefont {Turan}, \citenamefont {Taminiau},\ and\ \citenamefont {Hanson}}]{beukers_control_2024}%
  \BibitemOpen
  \bibfield  {author} {\bibinfo {author} {\bibfnamefont {H.~K.~C.}\ \bibnamefont {Beukers}}, \bibinfo {author} {\bibfnamefont {C.}~\bibnamefont {Waas}}, \bibinfo {author} {\bibfnamefont {M.}~\bibnamefont {Pasini}}, \bibinfo {author} {\bibfnamefont {H.~B.~v.}\ \bibnamefont {Ommen}}, \bibinfo {author} {\bibfnamefont {Z.}~\bibnamefont {Ademi}}, \bibinfo {author} {\bibfnamefont {M.}~\bibnamefont {Iuliano}}, \bibinfo {author} {\bibfnamefont {N.}~\bibnamefont {Codreanu}}, \bibinfo {author} {\bibfnamefont {J.~M.}\ \bibnamefont {Brevoord}}, \bibinfo {author} {\bibfnamefont {T.}~\bibnamefont {Turan}}, \bibinfo {author} {\bibfnamefont {T.~H.}\ \bibnamefont {Taminiau}}, \ and\ \bibinfo {author} {\bibfnamefont {R.}~\bibnamefont {Hanson}},\ }\href {\doibase 10.48550/arXiv.2409.08977} {\enquote {\bibinfo {title} {Control of solid-state nuclear spin qubits using an electron spin-1/2},}\ } (\bibinfo {year} {2024}),\ \bibinfo {note} {arXiv:2409.08977}\BibitemShut {NoStop}%
\bibitem [{\citenamefont {Rugar}\ \emph {et~al.}(2019)\citenamefont {Rugar}, \citenamefont {Dory}, \citenamefont {Sun},\ and\ \citenamefont {Vučković}}]{rugar_characterization_2019}%
  \BibitemOpen
  \bibfield  {author} {\bibinfo {author} {\bibfnamefont {A.~E.}\ \bibnamefont {Rugar}}, \bibinfo {author} {\bibfnamefont {C.}~\bibnamefont {Dory}}, \bibinfo {author} {\bibfnamefont {S.}~\bibnamefont {Sun}}, \ and\ \bibinfo {author} {\bibfnamefont {J.}~\bibnamefont {Vučković}},\ }\bibfield  {title} {{\selectlanguage {en}\enquote {\bibinfo {title} {Characterization of optical and spin properties of single tin-vacancy centers in diamond nanopillars},}\ }}\href {\doibase 10.1103/PhysRevB.99.205417} {\bibfield  {journal} {\bibinfo  {journal} {Phys. Rev. B}\ }\textbf {\bibinfo {volume} {99}},\ \bibinfo {pages} {205417} (\bibinfo {year} {2019})}\BibitemShut {NoStop}%
\bibitem [{\citenamefont {Pasini}\ \emph {et~al.}(2024)\citenamefont {Pasini}, \citenamefont {Codreanu}, \citenamefont {Turan}, \citenamefont {Riera~Moral}, \citenamefont {Primavera}, \citenamefont {De~Santis}, \citenamefont {Beukers}, \citenamefont {Brevoord}, \citenamefont {Waas}, \citenamefont {Borregaard},\ and\ \citenamefont {Hanson}}]{pasini_nonlinear_2024}%
  \BibitemOpen
  \bibfield  {author} {\bibinfo {author} {\bibfnamefont {M.}~\bibnamefont {Pasini}}, \bibinfo {author} {\bibfnamefont {N.}~\bibnamefont {Codreanu}}, \bibinfo {author} {\bibfnamefont {T.}~\bibnamefont {Turan}}, \bibinfo {author} {\bibfnamefont {A.}~\bibnamefont {Riera~Moral}}, \bibinfo {author} {\bibfnamefont {C.~F.}\ \bibnamefont {Primavera}}, \bibinfo {author} {\bibfnamefont {L.}~\bibnamefont {De~Santis}}, \bibinfo {author} {\bibfnamefont {H.~K.}\ \bibnamefont {Beukers}}, \bibinfo {author} {\bibfnamefont {J.~M.}\ \bibnamefont {Brevoord}}, \bibinfo {author} {\bibfnamefont {C.}~\bibnamefont {Waas}}, \bibinfo {author} {\bibfnamefont {J.}~\bibnamefont {Borregaard}}, \ and\ \bibinfo {author} {\bibfnamefont {R.}~\bibnamefont {Hanson}},\ }\bibfield  {title} {\enquote {\bibinfo {title} {Nonlinear {Quantum} {Photonics} with a {Tin}-{Vacancy} {Center} {Coupled} to a {One}-{Dimensional} {Diamond} {Waveguide}},}\ }\href {\doibase 10.1103/PhysRevLett.133.023603} {\bibfield  {journal} {\bibinfo  {journal} {Phys. Rev.
  Lett.}\ }\textbf {\bibinfo {volume} {133}},\ \bibinfo {pages} {023603} (\bibinfo {year} {2024})}\BibitemShut {NoStop}%
\bibitem [{\citenamefont {Clark}\ \emph {et~al.}(2024)\citenamefont {Clark}, \citenamefont {Raniwala}, \citenamefont {Koppa}, \citenamefont {Chen}, \citenamefont {Leenheer}, \citenamefont {Zimmermann}, \citenamefont {Dong}, \citenamefont {Li}, \citenamefont {Wen}, \citenamefont {Dominguez}, \citenamefont {Trusheim}, \citenamefont {Gilbert}, \citenamefont {Eichenfield},\ and\ \citenamefont {Englund}}]{clark_nanoelectromechanical_2024}%
  \BibitemOpen
  \bibfield  {author} {\bibinfo {author} {\bibfnamefont {G.}~\bibnamefont {Clark}}, \bibinfo {author} {\bibfnamefont {H.}~\bibnamefont {Raniwala}}, \bibinfo {author} {\bibfnamefont {M.}~\bibnamefont {Koppa}}, \bibinfo {author} {\bibfnamefont {K.}~\bibnamefont {Chen}}, \bibinfo {author} {\bibfnamefont {A.}~\bibnamefont {Leenheer}}, \bibinfo {author} {\bibfnamefont {M.}~\bibnamefont {Zimmermann}}, \bibinfo {author} {\bibfnamefont {M.}~\bibnamefont {Dong}}, \bibinfo {author} {\bibfnamefont {L.}~\bibnamefont {Li}}, \bibinfo {author} {\bibfnamefont {Y.~H.}\ \bibnamefont {Wen}}, \bibinfo {author} {\bibfnamefont {D.}~\bibnamefont {Dominguez}}, \bibinfo {author} {\bibfnamefont {M.}~\bibnamefont {Trusheim}}, \bibinfo {author} {\bibfnamefont {G.}~\bibnamefont {Gilbert}}, \bibinfo {author} {\bibfnamefont {M.}~\bibnamefont {Eichenfield}}, \ and\ \bibinfo {author} {\bibfnamefont {D.}~\bibnamefont {Englund}},\ }\bibfield  {title} {\enquote {\bibinfo {title} {Nanoelectromechanical {Control} of {Spin}–{Photon} {Interfaces}
  in a {Hybrid} {Quantum} {System} on {Chip}},}\ }\href {\doibase 10.1021/acs.nanolett.3c04301} {\bibfield  {journal} {\bibinfo  {journal} {Nano Letters}\ } (\bibinfo {year} {2024}),\ 10.1021/acs.nanolett.3c04301}\BibitemShut {NoStop}%
\bibitem [{\citenamefont {Arjona~Martínez}\ \emph {et~al.}(2022)\citenamefont {Arjona~Martínez}, \citenamefont {Parker}, \citenamefont {Chen}, \citenamefont {Purser}, \citenamefont {Li}, \citenamefont {Michaels}, \citenamefont {Stramma}, \citenamefont {Debroux}, \citenamefont {Harris}, \citenamefont {Hayhurst~Appel}, \citenamefont {Nichols}, \citenamefont {Trusheim}, \citenamefont {Gangloff}, \citenamefont {Englund},\ and\ \citenamefont {Atatüre}}]{arjona_martinez_photonic_2022}%
  \BibitemOpen
  \bibfield  {author} {\bibinfo {author} {\bibfnamefont {J.}~\bibnamefont {Arjona~Martínez}}, \bibinfo {author} {\bibfnamefont {R.~A.}\ \bibnamefont {Parker}}, \bibinfo {author} {\bibfnamefont {K.~C.}\ \bibnamefont {Chen}}, \bibinfo {author} {\bibfnamefont {C.~M.}\ \bibnamefont {Purser}}, \bibinfo {author} {\bibfnamefont {L.}~\bibnamefont {Li}}, \bibinfo {author} {\bibfnamefont {C.~P.}\ \bibnamefont {Michaels}}, \bibinfo {author} {\bibfnamefont {A.~M.}\ \bibnamefont {Stramma}}, \bibinfo {author} {\bibfnamefont {R.}~\bibnamefont {Debroux}}, \bibinfo {author} {\bibfnamefont {I.~B.}\ \bibnamefont {Harris}}, \bibinfo {author} {\bibfnamefont {M.}~\bibnamefont {Hayhurst~Appel}}, \bibinfo {author} {\bibfnamefont {E.~C.}\ \bibnamefont {Nichols}}, \bibinfo {author} {\bibfnamefont {M.~E.}\ \bibnamefont {Trusheim}}, \bibinfo {author} {\bibfnamefont {D.~A.}\ \bibnamefont {Gangloff}}, \bibinfo {author} {\bibfnamefont {D.}~\bibnamefont {Englund}}, \ and\ \bibinfo {author} {\bibfnamefont {M.}~\bibnamefont {Atatüre}},\
  }\bibfield  {title} {\enquote {\bibinfo {title} {Photonic {Indistinguishability} of the {Tin}-{Vacancy} {Center} in {Nanostructured} {Diamond}},}\ }\href {\doibase 10.1103/PhysRevLett.129.173603} {\bibfield  {journal} {\bibinfo  {journal} {Phys. Rev. Lett.}\ }\textbf {\bibinfo {volume} {129}},\ \bibinfo {pages} {173603} (\bibinfo {year} {2022})}\BibitemShut {NoStop}%
\bibitem [{\citenamefont {Bradac}\ \emph {et~al.}(2019)\citenamefont {Bradac}, \citenamefont {Gao}, \citenamefont {Forneris}, \citenamefont {Trusheim},\ and\ \citenamefont {Aharonovich}}]{bradac_quantum_2019}%
  \BibitemOpen
  \bibfield  {author} {\bibinfo {author} {\bibfnamefont {C.}~\bibnamefont {Bradac}}, \bibinfo {author} {\bibfnamefont {W.}~\bibnamefont {Gao}}, \bibinfo {author} {\bibfnamefont {J.}~\bibnamefont {Forneris}}, \bibinfo {author} {\bibfnamefont {M.~E.}\ \bibnamefont {Trusheim}}, \ and\ \bibinfo {author} {\bibfnamefont {I.}~\bibnamefont {Aharonovich}},\ }\bibfield  {title} {\enquote {\bibinfo {title} {Quantum nanophotonics with group {IV} defects in diamond},}\ }\href {\doibase 10.1038/s41467-019-13332-w} {\bibfield  {journal} {\bibinfo  {journal} {Nat. Commun.}\ }\textbf {\bibinfo {volume} {10}},\ \bibinfo {pages} {5625} (\bibinfo {year} {2019})}\BibitemShut {NoStop}%
\bibitem [{\citenamefont {Ruf}\ \emph {et~al.}(2021)\citenamefont {Ruf}, \citenamefont {Wan}, \citenamefont {Choi}, \citenamefont {Englund},\ and\ \citenamefont {Hanson}}]{ruf_quantum_2021}%
  \BibitemOpen
  \bibfield  {author} {\bibinfo {author} {\bibfnamefont {M.}~\bibnamefont {Ruf}}, \bibinfo {author} {\bibfnamefont {N.~H.}\ \bibnamefont {Wan}}, \bibinfo {author} {\bibfnamefont {H.}~\bibnamefont {Choi}}, \bibinfo {author} {\bibfnamefont {D.}~\bibnamefont {Englund}}, \ and\ \bibinfo {author} {\bibfnamefont {R.}~\bibnamefont {Hanson}},\ }\bibfield  {title} {\enquote {\bibinfo {title} {Quantum networks based on color centers in diamond},}\ }\href {\doibase 10.1063/5.0056534} {\bibfield  {journal} {\bibinfo  {journal} {Journal of Applied Physics}\ }\textbf {\bibinfo {volume} {130}},\ \bibinfo {pages} {070901} (\bibinfo {year} {2021})}\BibitemShut {NoStop}%
\bibitem [{\citenamefont {Choi}\ \emph {et~al.}(2019)\citenamefont {Choi}, \citenamefont {Pant}, \citenamefont {Guha},\ and\ \citenamefont {Englund}}]{choi_percolation-based_2019}%
  \BibitemOpen
  \bibfield  {author} {\bibinfo {author} {\bibfnamefont {H.}~\bibnamefont {Choi}}, \bibinfo {author} {\bibfnamefont {M.}~\bibnamefont {Pant}}, \bibinfo {author} {\bibfnamefont {S.}~\bibnamefont {Guha}}, \ and\ \bibinfo {author} {\bibfnamefont {D.}~\bibnamefont {Englund}},\ }\bibfield  {title} {\enquote {\bibinfo {title} {Percolation-based architecture for cluster state creation using photon-mediated entanglement between atomic memories},}\ }\href {\doibase 10.1038/s41534-019-0215-2} {\bibfield  {journal} {\bibinfo  {journal} {npj Quantum Inf}\ }\textbf {\bibinfo {volume} {5}},\ \bibinfo {pages} {1--7} (\bibinfo {year} {2019})}\BibitemShut {NoStop}%
\bibitem [{\citenamefont {Nickerson}, \citenamefont {Fitzsimons},\ and\ \citenamefont {Benjamin}(2014)}]{nickerson_freely_2014}%
  \BibitemOpen
  \bibfield  {author} {\bibinfo {author} {\bibfnamefont {N.~H.}\ \bibnamefont {Nickerson}}, \bibinfo {author} {\bibfnamefont {J.~F.}\ \bibnamefont {Fitzsimons}}, \ and\ \bibinfo {author} {\bibfnamefont {S.~C.}\ \bibnamefont {Benjamin}},\ }\bibfield  {title} {\enquote {\bibinfo {title} {Freely {Scalable} {Quantum} {Technologies} {Using} {Cells} of 5-to-50 {Qubits} with {Very} {Lossy} and {Noisy} {Photonic} {Links}},}\ }\href {\doibase 10.1103/PhysRevX.4.041041} {\bibfield  {journal} {\bibinfo  {journal} {Phys. Rev. X}\ }\textbf {\bibinfo {volume} {4}},\ \bibinfo {pages} {041041} (\bibinfo {year} {2014})}\BibitemShut {NoStop}%
\bibitem [{\citenamefont {Nemoto}\ \emph {et~al.}(2014)\citenamefont {Nemoto}, \citenamefont {Trupke}, \citenamefont {Devitt}, \citenamefont {Stephens}, \citenamefont {Scharfenberger}, \citenamefont {Buczak}, \citenamefont {Nöbauer}, \citenamefont {Everitt}, \citenamefont {Schmiedmayer},\ and\ \citenamefont {Munro}}]{nemoto_photonic_2014}%
  \BibitemOpen
  \bibfield  {author} {\bibinfo {author} {\bibfnamefont {K.}~\bibnamefont {Nemoto}}, \bibinfo {author} {\bibfnamefont {M.}~\bibnamefont {Trupke}}, \bibinfo {author} {\bibfnamefont {S.~J.}\ \bibnamefont {Devitt}}, \bibinfo {author} {\bibfnamefont {A.~M.}\ \bibnamefont {Stephens}}, \bibinfo {author} {\bibfnamefont {B.}~\bibnamefont {Scharfenberger}}, \bibinfo {author} {\bibfnamefont {K.}~\bibnamefont {Buczak}}, \bibinfo {author} {\bibfnamefont {T.}~\bibnamefont {Nöbauer}}, \bibinfo {author} {\bibfnamefont {M.~S.}\ \bibnamefont {Everitt}}, \bibinfo {author} {\bibfnamefont {J.}~\bibnamefont {Schmiedmayer}}, \ and\ \bibinfo {author} {\bibfnamefont {W.~J.}\ \bibnamefont {Munro}},\ }\bibfield  {title} {\enquote {\bibinfo {title} {Photonic {Architecture} for {Scalable} {Quantum} {Information} {Processing} in {Diamond}},}\ }\href {\doibase 10.1103/PhysRevX.4.031022} {\bibfield  {journal} {\bibinfo  {journal} {Phys. Rev. X}\ }\textbf {\bibinfo {volume} {4}},\ \bibinfo {pages} {031022} (\bibinfo {year}
  {2014})}\BibitemShut {NoStop}%
\bibitem [{\citenamefont {De~Santis}\ \emph {et~al.}(2021)\citenamefont {De~Santis}, \citenamefont {Trusheim}, \citenamefont {Chen},\ and\ \citenamefont {Englund}}]{de_santis_investigation_2021}%
  \BibitemOpen
  \bibfield  {author} {\bibinfo {author} {\bibfnamefont {L.}~\bibnamefont {De~Santis}}, \bibinfo {author} {\bibfnamefont {M.}~\bibnamefont {Trusheim}}, \bibinfo {author} {\bibfnamefont {K.}~\bibnamefont {Chen}}, \ and\ \bibinfo {author} {\bibfnamefont {D.}~\bibnamefont {Englund}},\ }\bibfield  {title} {\enquote {\bibinfo {title} {Investigation of the {Stark} {Effect} on a {Centrosymmetric} {Quantum} {Emitter} in {Diamond}},}\ }\href {\doibase 10.1103/PhysRevLett.127.147402} {\bibfield  {journal} {\bibinfo  {journal} {Phys. Rev. Lett.}\ }\textbf {\bibinfo {volume} {127}},\ \bibinfo {pages} {147402} (\bibinfo {year} {2021})}\BibitemShut {NoStop}%
\bibitem [{\citenamefont {Aghaeimeibodi}\ \emph {et~al.}(2021)\citenamefont {Aghaeimeibodi}, \citenamefont {Riedel}, \citenamefont {Rugar}, \citenamefont {Dory},\ and\ \citenamefont {Vučković}}]{aghaeimeibodi_electrical_2021}%
  \BibitemOpen
  \bibfield  {author} {\bibinfo {author} {\bibfnamefont {S.}~\bibnamefont {Aghaeimeibodi}}, \bibinfo {author} {\bibfnamefont {D.}~\bibnamefont {Riedel}}, \bibinfo {author} {\bibfnamefont {A.~E.}\ \bibnamefont {Rugar}}, \bibinfo {author} {\bibfnamefont {C.}~\bibnamefont {Dory}}, \ and\ \bibinfo {author} {\bibfnamefont {J.}~\bibnamefont {Vučković}},\ }\bibfield  {title} {\enquote {\bibinfo {title} {Electrical {Tuning} of {Tin}-{Vacancy} {Centers} in {Diamond}},}\ }\href {\doibase 10.1103/PhysRevApplied.15.064010} {\bibfield  {journal} {\bibinfo  {journal} {Phys. Rev. Appl.}\ }\textbf {\bibinfo {volume} {15}},\ \bibinfo {pages} {064010} (\bibinfo {year} {2021})}\BibitemShut {NoStop}%
\bibitem [{\citenamefont {Maity}\ \emph {et~al.}(2018)\citenamefont {Maity}, \citenamefont {Shao}, \citenamefont {Sohn}, \citenamefont {Meesala}, \citenamefont {Machielse}, \citenamefont {Bielejec}, \citenamefont {Markham},\ and\ \citenamefont {Lončar}}]{maity_spectral_2018}%
  \BibitemOpen
  \bibfield  {author} {\bibinfo {author} {\bibfnamefont {S.}~\bibnamefont {Maity}}, \bibinfo {author} {\bibfnamefont {L.}~\bibnamefont {Shao}}, \bibinfo {author} {\bibfnamefont {Y.-I.}\ \bibnamefont {Sohn}}, \bibinfo {author} {\bibfnamefont {S.}~\bibnamefont {Meesala}}, \bibinfo {author} {\bibfnamefont {B.}~\bibnamefont {Machielse}}, \bibinfo {author} {\bibfnamefont {E.}~\bibnamefont {Bielejec}}, \bibinfo {author} {\bibfnamefont {M.}~\bibnamefont {Markham}}, \ and\ \bibinfo {author} {\bibfnamefont {M.}~\bibnamefont {Lončar}},\ }\bibfield  {title} {\enquote {\bibinfo {title} {Spectral {Alignment} of {Single}-{Photon} {Emitters} in {Diamond} using {Strain} {Gradient}},}\ }\href {\doibase 10.1103/PhysRevApplied.10.024050} {\bibfield  {journal} {\bibinfo  {journal} {Phys. Rev. Appl.}\ }\textbf {\bibinfo {volume} {10}},\ \bibinfo {pages} {024050} (\bibinfo {year} {2018})}\BibitemShut {NoStop}%
\bibitem [{\citenamefont {Meesala}\ \emph {et~al.}(2018)\citenamefont {Meesala}, \citenamefont {Sohn}, \citenamefont {Pingault}, \citenamefont {Shao}, \citenamefont {Atikian}, \citenamefont {Holzgrafe}, \citenamefont {Gundogan}, \citenamefont {Stavrakas}, \citenamefont {Sipahigil}, \citenamefont {Chia}, \citenamefont {Burek}, \citenamefont {Zhang}, \citenamefont {Wu}, \citenamefont {Pacheco}, \citenamefont {Abraham}, \citenamefont {Bielejec}, \citenamefont {Lukin}, \citenamefont {Atature},\ and\ \citenamefont {Loncar}}]{meesala_strain_2018}%
  \BibitemOpen
  \bibfield  {author} {\bibinfo {author} {\bibfnamefont {S.}~\bibnamefont {Meesala}}, \bibinfo {author} {\bibfnamefont {Y.-I.}\ \bibnamefont {Sohn}}, \bibinfo {author} {\bibfnamefont {B.}~\bibnamefont {Pingault}}, \bibinfo {author} {\bibfnamefont {L.}~\bibnamefont {Shao}}, \bibinfo {author} {\bibfnamefont {H.~A.}\ \bibnamefont {Atikian}}, \bibinfo {author} {\bibfnamefont {J.}~\bibnamefont {Holzgrafe}}, \bibinfo {author} {\bibfnamefont {M.}~\bibnamefont {Gundogan}}, \bibinfo {author} {\bibfnamefont {C.}~\bibnamefont {Stavrakas}}, \bibinfo {author} {\bibfnamefont {A.}~\bibnamefont {Sipahigil}}, \bibinfo {author} {\bibfnamefont {C.}~\bibnamefont {Chia}}, \bibinfo {author} {\bibfnamefont {M.~J.}\ \bibnamefont {Burek}}, \bibinfo {author} {\bibfnamefont {M.}~\bibnamefont {Zhang}}, \bibinfo {author} {\bibfnamefont {L.}~\bibnamefont {Wu}}, \bibinfo {author} {\bibfnamefont {J.~L.}\ \bibnamefont {Pacheco}}, \bibinfo {author} {\bibfnamefont {J.}~\bibnamefont {Abraham}}, \bibinfo {author} {\bibfnamefont {E.}~\bibnamefont
  {Bielejec}}, \bibinfo {author} {\bibfnamefont {M.~D.}\ \bibnamefont {Lukin}}, \bibinfo {author} {\bibfnamefont {M.}~\bibnamefont {Atature}}, \ and\ \bibinfo {author} {\bibfnamefont {M.}~\bibnamefont {Loncar}},\ }\bibfield  {title} {\enquote {\bibinfo {title} {Strain engineering of the silicon-vacancy center in diamond},}\ }\href {\doibase 10.1103/PhysRevB.97.205444} {\bibfield  {journal} {\bibinfo  {journal} {Phys. Rev. B}\ }\textbf {\bibinfo {volume} {97}},\ \bibinfo {pages} {205444} (\bibinfo {year} {2018})}\BibitemShut {NoStop}%
\bibitem [{\citenamefont {Sohn}\ \emph {et~al.}(2018)\citenamefont {Sohn}, \citenamefont {Meesala}, \citenamefont {Pingault}, \citenamefont {Atikian}, \citenamefont {Holzgrafe}, \citenamefont {Gündoğan}, \citenamefont {Stavrakas}, \citenamefont {Stanley}, \citenamefont {Sipahigil}, \citenamefont {Choi}, \citenamefont {Zhang}, \citenamefont {Pacheco}, \citenamefont {Abraham}, \citenamefont {Bielejec}, \citenamefont {Lukin}, \citenamefont {Atatüre},\ and\ \citenamefont {Lončar}}]{sohn_controlling_2018}%
  \BibitemOpen
  \bibfield  {author} {\bibinfo {author} {\bibfnamefont {Y.-I.}\ \bibnamefont {Sohn}}, \bibinfo {author} {\bibfnamefont {S.}~\bibnamefont {Meesala}}, \bibinfo {author} {\bibfnamefont {B.}~\bibnamefont {Pingault}}, \bibinfo {author} {\bibfnamefont {H.~A.}\ \bibnamefont {Atikian}}, \bibinfo {author} {\bibfnamefont {J.}~\bibnamefont {Holzgrafe}}, \bibinfo {author} {\bibfnamefont {M.}~\bibnamefont {Gündoğan}}, \bibinfo {author} {\bibfnamefont {C.}~\bibnamefont {Stavrakas}}, \bibinfo {author} {\bibfnamefont {M.~J.}\ \bibnamefont {Stanley}}, \bibinfo {author} {\bibfnamefont {A.}~\bibnamefont {Sipahigil}}, \bibinfo {author} {\bibfnamefont {J.}~\bibnamefont {Choi}}, \bibinfo {author} {\bibfnamefont {M.}~\bibnamefont {Zhang}}, \bibinfo {author} {\bibfnamefont {J.~L.}\ \bibnamefont {Pacheco}}, \bibinfo {author} {\bibfnamefont {J.}~\bibnamefont {Abraham}}, \bibinfo {author} {\bibfnamefont {E.}~\bibnamefont {Bielejec}}, \bibinfo {author} {\bibfnamefont {M.~D.}\ \bibnamefont {Lukin}}, \bibinfo {author} {\bibfnamefont
  {M.}~\bibnamefont {Atatüre}}, \ and\ \bibinfo {author} {\bibfnamefont {M.}~\bibnamefont {Lončar}},\ }\bibfield  {title} {{\selectlanguage {en}\enquote {\bibinfo {title} {Controlling the coherence of a diamond spin qubit through its strain environment},}\ }}\href {\doibase 10.1038/s41467-018-04340-3} {\bibfield  {journal} {\bibinfo  {journal} {Nat Commun}\ }\textbf {\bibinfo {volume} {9}},\ \bibinfo {pages} {2012} (\bibinfo {year} {2018})}\BibitemShut {NoStop}%
\bibitem [{\citenamefont {Machielse}\ \emph {et~al.}(2019)\citenamefont {Machielse}, \citenamefont {Bogdanovic}, \citenamefont {Meesala}, \citenamefont {Gauthier}, \citenamefont {Burek}, \citenamefont {Joe}, \citenamefont {Chalupnik}, \citenamefont {Sohn}, \citenamefont {Holzgrafe}, \citenamefont {Evans}, \citenamefont {Chia}, \citenamefont {Atikian}, \citenamefont {Bhaskar}, \citenamefont {Sukachev}, \citenamefont {Shao}, \citenamefont {Maity}, \citenamefont {Lukin},\ and\ \citenamefont {Lončar}}]{machielse_quantum_2019}%
  \BibitemOpen
  \bibfield  {author} {\bibinfo {author} {\bibfnamefont {B.}~\bibnamefont {Machielse}}, \bibinfo {author} {\bibfnamefont {S.}~\bibnamefont {Bogdanovic}}, \bibinfo {author} {\bibfnamefont {S.}~\bibnamefont {Meesala}}, \bibinfo {author} {\bibfnamefont {S.}~\bibnamefont {Gauthier}}, \bibinfo {author} {\bibfnamefont {M.}~\bibnamefont {Burek}}, \bibinfo {author} {\bibfnamefont {G.}~\bibnamefont {Joe}}, \bibinfo {author} {\bibfnamefont {M.}~\bibnamefont {Chalupnik}}, \bibinfo {author} {\bibfnamefont {Y.}~\bibnamefont {Sohn}}, \bibinfo {author} {\bibfnamefont {J.}~\bibnamefont {Holzgrafe}}, \bibinfo {author} {\bibfnamefont {R.}~\bibnamefont {Evans}}, \bibinfo {author} {\bibfnamefont {C.}~\bibnamefont {Chia}}, \bibinfo {author} {\bibfnamefont {H.}~\bibnamefont {Atikian}}, \bibinfo {author} {\bibfnamefont {M.}~\bibnamefont {Bhaskar}}, \bibinfo {author} {\bibfnamefont {D.}~\bibnamefont {Sukachev}}, \bibinfo {author} {\bibfnamefont {L.}~\bibnamefont {Shao}}, \bibinfo {author} {\bibfnamefont {S.}~\bibnamefont {Maity}},
  \bibinfo {author} {\bibfnamefont {M.}~\bibnamefont {Lukin}}, \ and\ \bibinfo {author} {\bibfnamefont {M.}~\bibnamefont {Lončar}},\ }\bibfield  {title} {\enquote {\bibinfo {title} {Quantum {Interference} of {Electromechanically} {Stabilized} {Emitters} in {Nanophotonic} {Devices}},}\ }\href {\doibase 10.1103/PhysRevX.9.031022} {\bibfield  {journal} {\bibinfo  {journal} {Phys. Rev. X}\ }\textbf {\bibinfo {volume} {9}},\ \bibinfo {pages} {031022} (\bibinfo {year} {2019})}\BibitemShut {NoStop}%
\bibitem [{\citenamefont {Hepp}\ \emph {et~al.}(2014)\citenamefont {Hepp}, \citenamefont {Müller}, \citenamefont {Waselowski}, \citenamefont {Becker}, \citenamefont {Pingault}, \citenamefont {Sternschulte}, \citenamefont {Steinmüller-Nethl}, \citenamefont {Gali}, \citenamefont {Maze}, \citenamefont {Atatüre},\ and\ \citenamefont {Becher}}]{hepp_electronic_2014}%
  \BibitemOpen
  \bibfield  {author} {\bibinfo {author} {\bibfnamefont {C.}~\bibnamefont {Hepp}}, \bibinfo {author} {\bibfnamefont {T.}~\bibnamefont {Müller}}, \bibinfo {author} {\bibfnamefont {V.}~\bibnamefont {Waselowski}}, \bibinfo {author} {\bibfnamefont {J.~N.}\ \bibnamefont {Becker}}, \bibinfo {author} {\bibfnamefont {B.}~\bibnamefont {Pingault}}, \bibinfo {author} {\bibfnamefont {H.}~\bibnamefont {Sternschulte}}, \bibinfo {author} {\bibfnamefont {D.}~\bibnamefont {Steinmüller-Nethl}}, \bibinfo {author} {\bibfnamefont {A.}~\bibnamefont {Gali}}, \bibinfo {author} {\bibfnamefont {J.~R.}\ \bibnamefont {Maze}}, \bibinfo {author} {\bibfnamefont {M.}~\bibnamefont {Atatüre}}, \ and\ \bibinfo {author} {\bibfnamefont {C.}~\bibnamefont {Becher}},\ }\bibfield  {title} {\enquote {\bibinfo {title} {Electronic {Structure} of the {Silicon} {Vacancy} {Color} {Center} in {Diamond}},}\ }\href {\doibase 10.1103/PhysRevLett.112.036405} {\bibfield  {journal} {\bibinfo  {journal} {Phys. Rev. Lett.}\ }\textbf {\bibinfo {volume} {112}},\
  \bibinfo {pages} {036405} (\bibinfo {year} {2014})}\BibitemShut {NoStop}%
\bibitem [{\citenamefont {Pieplow}\ \emph {et~al.}(2024)\citenamefont {Pieplow}, \citenamefont {Torun}, \citenamefont {Munns}, \citenamefont {Herrmann}, \citenamefont {Thies}, \citenamefont {Pregnolato},\ and\ \citenamefont {Schröder}}]{pieplow_quantum_2024}%
  \BibitemOpen
  \bibfield  {author} {\bibinfo {author} {\bibfnamefont {G.}~\bibnamefont {Pieplow}}, \bibinfo {author} {\bibfnamefont {C.~G.}\ \bibnamefont {Torun}}, \bibinfo {author} {\bibfnamefont {J.~H.~D.}\ \bibnamefont {Munns}}, \bibinfo {author} {\bibfnamefont {F.~M.}\ \bibnamefont {Herrmann}}, \bibinfo {author} {\bibfnamefont {A.}~\bibnamefont {Thies}}, \bibinfo {author} {\bibfnamefont {T.}~\bibnamefont {Pregnolato}}, \ and\ \bibinfo {author} {\bibfnamefont {T.}~\bibnamefont {Schröder}},\ }\href {\doibase 10.48550/arXiv.2401.14290} {\enquote {\bibinfo {title} {Quantum {Electrometer} for {Time}-{Resolved} {Material} {Science} at the {Atomic} {Lattice} {Scale}},}\ } (\bibinfo {year} {2024}),\ \bibinfo {note} {arXiv:2401.14290}\BibitemShut {NoStop}%
\bibitem [{\citenamefont {Herrmann}\ \emph {et~al.}(2024)\citenamefont {Herrmann}, \citenamefont {Fischer}, \citenamefont {Brevoord}, \citenamefont {Sauerzapf}, \citenamefont {Wienhoven}, \citenamefont {Feije}, \citenamefont {Pasini}, \citenamefont {Eschen}, \citenamefont {Ruf}, \citenamefont {Weaver},\ and\ \citenamefont {Hanson}}]{herrmann_coherent_2024}%
  \BibitemOpen
  \bibfield  {author} {\bibinfo {author} {\bibfnamefont {Y.}~\bibnamefont {Herrmann}}, \bibinfo {author} {\bibfnamefont {J.}~\bibnamefont {Fischer}}, \bibinfo {author} {\bibfnamefont {J.~M.}\ \bibnamefont {Brevoord}}, \bibinfo {author} {\bibfnamefont {C.}~\bibnamefont {Sauerzapf}}, \bibinfo {author} {\bibfnamefont {L.~G.}\ \bibnamefont {Wienhoven}}, \bibinfo {author} {\bibfnamefont {L.~J.}\ \bibnamefont {Feije}}, \bibinfo {author} {\bibfnamefont {M.}~\bibnamefont {Pasini}}, \bibinfo {author} {\bibfnamefont {M.}~\bibnamefont {Eschen}}, \bibinfo {author} {\bibfnamefont {M.}~\bibnamefont {Ruf}}, \bibinfo {author} {\bibfnamefont {M.~J.}\ \bibnamefont {Weaver}}, \ and\ \bibinfo {author} {\bibfnamefont {R.}~\bibnamefont {Hanson}},\ }\bibfield  {title} {\enquote {\bibinfo {title} {Coherent {Coupling} of a {Diamond} {Tin}-{Vacancy} {Center} to a {Tunable} {Open} {Microcavity}},}\ }\href {\doibase 10.1103/PhysRevX.14.041013} {\bibfield  {journal} {\bibinfo  {journal} {Phys. Rev. X}\ }\textbf {\bibinfo {volume} {14}},\
  \bibinfo {pages} {041013} (\bibinfo {year} {2024})}\BibitemShut {NoStop}%
\bibitem [{\citenamefont {Li}\ \emph {et~al.}(2024{\natexlab{b}})\citenamefont {Li}, \citenamefont {Guo}, \citenamefont {Jin}, \citenamefont {Andreoli}, \citenamefont {Bilgin}, \citenamefont {Awschalom}, \citenamefont {Delegan}, \citenamefont {Heremans}, \citenamefont {Chang}, \citenamefont {Galli},\ and\ \citenamefont {High}}]{li_atomic_2024}%
  \BibitemOpen
  \bibfield  {author} {\bibinfo {author} {\bibfnamefont {Z.}~\bibnamefont {Li}}, \bibinfo {author} {\bibfnamefont {X.}~\bibnamefont {Guo}}, \bibinfo {author} {\bibfnamefont {Y.}~\bibnamefont {Jin}}, \bibinfo {author} {\bibfnamefont {F.}~\bibnamefont {Andreoli}}, \bibinfo {author} {\bibfnamefont {A.}~\bibnamefont {Bilgin}}, \bibinfo {author} {\bibfnamefont {D.~D.}\ \bibnamefont {Awschalom}}, \bibinfo {author} {\bibfnamefont {N.}~\bibnamefont {Delegan}}, \bibinfo {author} {\bibfnamefont {F.~J.}\ \bibnamefont {Heremans}}, \bibinfo {author} {\bibfnamefont {D.}~\bibnamefont {Chang}}, \bibinfo {author} {\bibfnamefont {G.}~\bibnamefont {Galli}}, \ and\ \bibinfo {author} {\bibfnamefont {A.~A.}\ \bibnamefont {High}},\ }\bibfield  {title} {{\selectlanguage {en}\enquote {\bibinfo {title} {Atomic optical antennas in solids},}\ }}\href {\doibase 10.1038/s41566-024-01456-5} {\bibfield  {journal} {\bibinfo  {journal} {Nat. Photon.}\ } (\bibinfo {year} {2024}{\natexlab{b}}),\ 10.1038/s41566-024-01456-5}\BibitemShut {NoStop}%
\bibitem [{\citenamefont {Ziegler}, \citenamefont {Ziegler},\ and\ \citenamefont {Biersack}(2010)}]{ziegler_srim_2010}%
  \BibitemOpen
  \bibfield  {author} {\bibinfo {author} {\bibfnamefont {J.~F.}\ \bibnamefont {Ziegler}}, \bibinfo {author} {\bibfnamefont {M.}~\bibnamefont {Ziegler}}, \ and\ \bibinfo {author} {\bibfnamefont {J.}~\bibnamefont {Biersack}},\ }\bibfield  {title} {\enquote {\bibinfo {title} {{SRIM} – {The} stopping and range of ions in matter (2010)},}\ }\href {\doibase 10.1016/j.nimb.2010.02.091} {\bibfield  {journal} {\bibinfo  {journal} {Nuclear Instruments and Methods in Physics Research Section B: Beam Interactions with Materials and Atoms}\ }\textbf {\bibinfo {volume} {268}},\ \bibinfo {pages} {1818--1823} (\bibinfo {year} {2010})}\BibitemShut {NoStop}%
\bibitem [{noa()}]{noauthor_ansys_nodate}%
  \BibitemOpen
  \href@noop {} {\enquote {\bibinfo {title} {Ansys® {Academic} {Research} {Mechanical}, {Release} 18.1},}\ }\BibitemShut {NoStop}%
\bibitem [{\citenamefont {Bernien}\ \emph {et~al.}(2013)\citenamefont {Bernien}, \citenamefont {Hensen}, \citenamefont {Pfaff}, \citenamefont {Koolstra}, \citenamefont {Blok}, \citenamefont {Robledo}, \citenamefont {Taminiau}, \citenamefont {Markham}, \citenamefont {Twitchen}, \citenamefont {Childress},\ and\ \citenamefont {Hanson}}]{bernien_heralded_2013}%
  \BibitemOpen
  \bibfield  {author} {\bibinfo {author} {\bibfnamefont {H.}~\bibnamefont {Bernien}}, \bibinfo {author} {\bibfnamefont {B.}~\bibnamefont {Hensen}}, \bibinfo {author} {\bibfnamefont {W.}~\bibnamefont {Pfaff}}, \bibinfo {author} {\bibfnamefont {G.}~\bibnamefont {Koolstra}}, \bibinfo {author} {\bibfnamefont {M.~S.}\ \bibnamefont {Blok}}, \bibinfo {author} {\bibfnamefont {L.}~\bibnamefont {Robledo}}, \bibinfo {author} {\bibfnamefont {T.~H.}\ \bibnamefont {Taminiau}}, \bibinfo {author} {\bibfnamefont {M.}~\bibnamefont {Markham}}, \bibinfo {author} {\bibfnamefont {D.~J.}\ \bibnamefont {Twitchen}}, \bibinfo {author} {\bibfnamefont {L.}~\bibnamefont {Childress}}, \ and\ \bibinfo {author} {\bibfnamefont {R.}~\bibnamefont {Hanson}},\ }\bibfield  {title} {\enquote {\bibinfo {title} {Heralded entanglement between solid-state qubits separated by three metres},}\ }\href {\doibase 10.1038/nature12016} {\bibfield  {journal} {\bibinfo  {journal} {Nature}\ }\textbf {\bibinfo {volume} {497}},\ \bibinfo {pages} {86--90}
  (\bibinfo {year} {2013})}\BibitemShut {NoStop}%
\bibitem [{\citenamefont {Brevoord}\ \emph {et~al.}(2024{\natexlab{b}})\citenamefont {Brevoord}, \citenamefont {Wienhoven}, \citenamefont {Codreanu}, \citenamefont {Ishiguro}, \citenamefont {van Leeuwen}, \citenamefont {Iuliano}, \citenamefont {De~Santis}, \citenamefont {Waas}, \citenamefont {Beukers}, \citenamefont {Turan}, \citenamefont {Errando-Herranz}, \citenamefont {Kawaguchi},\ and\ \citenamefont {Hanson}}]{brevoord_data_2024}%
  \BibitemOpen
  \bibfield  {author} {\bibinfo {author} {\bibfnamefont {J.~M.}\ \bibnamefont {Brevoord}}, \bibinfo {author} {\bibfnamefont {L.}~\bibnamefont {Wienhoven}}, \bibinfo {author} {\bibfnamefont {N.}~\bibnamefont {Codreanu}}, \bibinfo {author} {\bibfnamefont {T.}~\bibnamefont {Ishiguro}}, \bibinfo {author} {\bibfnamefont {E.}~\bibnamefont {van Leeuwen}}, \bibinfo {author} {\bibfnamefont {M.}~\bibnamefont {Iuliano}}, \bibinfo {author} {\bibfnamefont {L.}~\bibnamefont {De~Santis}}, \bibinfo {author} {\bibfnamefont {C.}~\bibnamefont {Waas}}, \bibinfo {author} {\bibfnamefont {H.~K.~C.}\ \bibnamefont {Beukers}}, \bibinfo {author} {\bibfnamefont {T.~L.}\ \bibnamefont {Turan}}, \bibinfo {author} {\bibfnamefont {C.}~\bibnamefont {Errando-Herranz}}, \bibinfo {author} {\bibfnamefont {K.}~\bibnamefont {Kawaguchi}}, \ and\ \bibinfo {author} {\bibfnamefont {R.}~\bibnamefont {Hanson}},\ }\href {\doibase 10.4121/6b89815e-d73c-4abd-93cc-a0514c2780ae} {{\selectlanguage {english}\enquote {\bibinfo {title} {Data underlying the
  publication "{Large}-{Range} {Tuning} and {Stabilization} of the {Optical} {Transition} of {Diamond} {Tin}-{Vacancy} {Centers} by {In}-{Situ} {Strain} {Control}"},}\ }} (\bibinfo {year} {2024}{\natexlab{b}})\BibitemShut {NoStop}%
\bibitem [{\citenamefont {Khanaliloo}\ \emph {et~al.}(2015)\citenamefont {Khanaliloo}, \citenamefont {Mitchell}, \citenamefont {Hryciw},\ and\ \citenamefont {Barclay}}]{khanaliloo_high-qv_2015}%
  \BibitemOpen
  \bibfield  {author} {\bibinfo {author} {\bibfnamefont {B.}~\bibnamefont {Khanaliloo}}, \bibinfo {author} {\bibfnamefont {M.}~\bibnamefont {Mitchell}}, \bibinfo {author} {\bibfnamefont {A.~C.}\ \bibnamefont {Hryciw}}, \ and\ \bibinfo {author} {\bibfnamefont {P.~E.}\ \bibnamefont {Barclay}},\ }\bibfield  {title} {\enquote {\bibinfo {title} {High-{Q}/{V} {Monolithic} {Diamond} {Microdisks} {Fabricated} with {Quasi}-isotropic {Etching}},}\ }\href {\doibase 10.1021/acs.nanolett.5b01346} {\bibfield  {journal} {\bibinfo  {journal} {Nano Lett.}\ }\textbf {\bibinfo {volume} {15}},\ \bibinfo {pages} {5131--5136} (\bibinfo {year} {2015})}\BibitemShut {NoStop}%
\bibitem [{\citenamefont {Mitchell}, \citenamefont {Lake},\ and\ \citenamefont {Barclay}(2019)}]{mitchell_realizing_2019}%
  \BibitemOpen
  \bibfield  {author} {\bibinfo {author} {\bibfnamefont {M.}~\bibnamefont {Mitchell}}, \bibinfo {author} {\bibfnamefont {D.~P.}\ \bibnamefont {Lake}}, \ and\ \bibinfo {author} {\bibfnamefont {P.~E.}\ \bibnamefont {Barclay}},\ }\bibfield  {title} {\enquote {\bibinfo {title} {Realizing {Q} {\textgreater} 300 000 in diamond microdisks for optomechanics via etch optimization},}\ }\href {\doibase 10.1063/1.5053122} {\bibfield  {journal} {\bibinfo  {journal} {APL Photonics}\ }\textbf {\bibinfo {volume} {4}} (\bibinfo {year} {2019}),\ 10.1063/1.5053122}\BibitemShut {NoStop}%
\bibitem [{\citenamefont {Mouradian}\ \emph {et~al.}(2017)\citenamefont {Mouradian}, \citenamefont {Wan}, \citenamefont {Schröder},\ and\ \citenamefont {Englund}}]{mouradian_rectangular_2017}%
  \BibitemOpen
  \bibfield  {author} {\bibinfo {author} {\bibfnamefont {S.}~\bibnamefont {Mouradian}}, \bibinfo {author} {\bibfnamefont {N.~H.}\ \bibnamefont {Wan}}, \bibinfo {author} {\bibfnamefont {T.}~\bibnamefont {Schröder}}, \ and\ \bibinfo {author} {\bibfnamefont {D.}~\bibnamefont {Englund}},\ }\bibfield  {title} {\enquote {\bibinfo {title} {Rectangular photonic crystal nanobeam cavities in bulk diamond},}\ }\href {\doibase 10.1063/1.4992118} {\bibfield  {journal} {\bibinfo  {journal} {Applied Physics Letters}\ }\textbf {\bibinfo {volume} {111}},\ \bibinfo {pages} {021103} (\bibinfo {year} {2017})}\BibitemShut {NoStop}%
\bibitem [{\citenamefont {Raa}\ \emph {et~al.}(2023)\citenamefont {Raa}, \citenamefont {Ervasti}, \citenamefont {Botma}, \citenamefont {Visser}, \citenamefont {Budhrani}, \citenamefont {van Rantwijk}, \citenamefont {Cadot}, \citenamefont {Vermeltfoort}, \citenamefont {Pompili}, \citenamefont {Stolk}, \citenamefont {Weaver}, \citenamefont {van~der Enden}, \citenamefont {de~Leeuw~Duarte}, \citenamefont {Teng}, \citenamefont {van Zwieten},\ and\ \citenamefont {Grooteman}}]{raa_qmi_2023}%
  \BibitemOpen
  \bibfield  {author} {\bibinfo {author} {\bibfnamefont {I.~T.}\ \bibnamefont {Raa}}, \bibinfo {author} {\bibfnamefont {H.~K.}\ \bibnamefont {Ervasti}}, \bibinfo {author} {\bibfnamefont {P.~J.}\ \bibnamefont {Botma}}, \bibinfo {author} {\bibfnamefont {L.~C.}\ \bibnamefont {Visser}}, \bibinfo {author} {\bibfnamefont {R.}~\bibnamefont {Budhrani}}, \bibinfo {author} {\bibfnamefont {J.~F.}\ \bibnamefont {van Rantwijk}}, \bibinfo {author} {\bibfnamefont {S.~P.}\ \bibnamefont {Cadot}}, \bibinfo {author} {\bibfnamefont {J.}~\bibnamefont {Vermeltfoort}}, \bibinfo {author} {\bibfnamefont {M.}~\bibnamefont {Pompili}}, \bibinfo {author} {\bibfnamefont {A.~J.}\ \bibnamefont {Stolk}}, \bibinfo {author} {\bibfnamefont {M.~J.}\ \bibnamefont {Weaver}}, \bibinfo {author} {\bibfnamefont {K.~L.}\ \bibnamefont {van~der Enden}}, \bibinfo {author} {\bibfnamefont {D.}~\bibnamefont {de~Leeuw~Duarte}}, \bibinfo {author} {\bibfnamefont {M.}~\bibnamefont {Teng}}, \bibinfo {author} {\bibfnamefont {J.}~\bibnamefont {van Zwieten}}, \ and\
  \bibinfo {author} {\bibfnamefont {F.}~\bibnamefont {Grooteman}},\ }\href {\doibase 10.4121/6D39C6DB-2F50-4A49-AD60-5BB08F40CB52} {\enquote {\bibinfo {title} {{QMI} - {Quantum} {Measurement} {Infrastructure}, a {Python} 3 framework for controlling laboratory equipment},}\ } (\bibinfo {year} {2023})\BibitemShut {NoStop}%
\bibitem [{\citenamefont {Dias}\ and\ \citenamefont {Mattos}(2015)}]{dias_uranium-zirconium_2015}%
  \BibitemOpen
  \bibfield  {author} {\bibinfo {author} {\bibfnamefont {M.~S.}\ \bibnamefont {Dias}}\ and\ \bibinfo {author} {\bibfnamefont {J.~R. L.~d.}\ \bibnamefont {Mattos}},\ }\bibfield  {title} {\enquote {\bibinfo {title} {Uranium-zirconium based alloys part {I}: reference points for thermophysical properties},}\ \ }(\bibinfo {address} {Brazil},\ \bibinfo {year} {2015})\BibitemShut {NoStop}%
\bibitem [{\citenamefont {Jacobson}\ and\ \citenamefont {Stoupin}(2019)}]{jacobson_thermal_2019}%
  \BibitemOpen
  \bibfield  {author} {\bibinfo {author} {\bibfnamefont {P.}~\bibnamefont {Jacobson}}\ and\ \bibinfo {author} {\bibfnamefont {S.}~\bibnamefont {Stoupin}},\ }\bibfield  {title} {\enquote {\bibinfo {title} {Thermal expansion coefficient of diamond in a wide temperature range},}\ }\href {\doibase 10.1016/j.diamond.2019.107469} {\bibfield  {journal} {\bibinfo  {journal} {Diamond and Related Materials}\ }\textbf {\bibinfo {volume} {97}},\ \bibinfo {pages} {107469} (\bibinfo {year} {2019})}\BibitemShut {NoStop}%
\bibitem [{\citenamefont {Klein}\ and\ \citenamefont {Cardinale}(1993)}]{klein_youngs_1993}%
  \BibitemOpen
  \bibfield  {author} {\bibinfo {author} {\bibfnamefont {C.~A.}\ \bibnamefont {Klein}}\ and\ \bibinfo {author} {\bibfnamefont {G.~F.}\ \bibnamefont {Cardinale}},\ }\bibfield  {title} {\enquote {\bibinfo {title} {Young's modulus and {Poisson}'s ratio of {CVD} diamond},}\ }\href {\doibase 10.1016/0925-9635(93)90250-6} {\bibfield  {journal} {\bibinfo  {journal} {Diamond and Related Materials}\ }\textbf {\bibinfo {volume} {2}},\ \bibinfo {pages} {918--923} (\bibinfo {year} {1993})}\BibitemShut {NoStop}%
\bibitem [{\citenamefont {Shabalin}(2014)}]{shabalin_ultra-high_2014}%
  \BibitemOpen
  \bibfield  {author} {\bibinfo {author} {\bibfnamefont {I.~L.}\ \bibnamefont {Shabalin}},\ }\href {\doibase 10.1007/978-94-007-7587-9} {\emph {\bibinfo {title} {Ultra-{High} {Temperature} {Materials} {I}: {Carbon} ({Graphene}/{Graphite}) and {Refractory} {Metals}}}}\ (\bibinfo  {publisher} {Springer Netherlands},\ \bibinfo {address} {Dordrecht},\ \bibinfo {year} {2014})\BibitemShut {NoStop}%
\end{thebibliography}%

\clearpage

\widetext
\begin{center}
\textbf{\large Supplementary Information: Large-Range Tuning and Stabilization of the Optical Transition of Diamond Tin-Vacancy Centers by In-Situ Strain Control.}
\end{center}
\setcounter{figure}{0}
\setcounter{page}{1}
\setcounter{section}{0}
\makeatletter
\renewcommand{\theequation}{S\arabic{equation}}
\renewcommand{\thefigure}{S\arabic{figure}}
\thispagestyle{empty}
\section{Device Fabrication}
The fabrication of the diamond MEMS devices is divided into three main stages: incorporating coherent SnVs$^-$ in diamond, fabricating diamond waveguides, and defining thin-film electrodes onto the diamond nanostructures.

The SnV incorporation process starts from a $\braket{001}$ surface-oriented electronic-grade single-crystal diamond (supplied by Element 6) and follows the same general process as described in~\cite{pasini_nonlinear_2024}. After an inorganic surface pre-clean and an inductively-coupled-plasma reactive-ion-etching (ICP-RIE) strain relieve etch, the diamond is implanted with $^{120}$Sn$^+$ ions at an acceleration voltage of \unit[350]{keV} and an implantation dose of $10^{11}$ ions/cm$^2$. Following the $^{120}$Sn$^+$ implantation, the diamond is cleaned in a boiling tri-acid solution (ratio 1:1:1 of HNO$_3$ (65$\%$):HClO$_4$ (70$\%$):H$_2$SO$_4$ (95$\%$)) and the SnVs are activated via vacuum-annealing. The graphitized diamond surface post-vacuum-annealing is removed with a combination of a boiling tri-acid clean and a short ICP-RIE \ch{O2} etch.

The diamond waveguide fabrication is based on the crystallographic-plane-dependent quasi-isotropic etch undercut method, developed in~\cite{pasini_nonlinear_2024, wan_large-scale_2020,ruf_quantum_2021,khanaliloo_high-qv_2015, mitchell_realizing_2019, mouradian_rectangular_2017}. We again follow the same general process as described in\cite{pasini_nonlinear_2024}, but here, the quasi-isotropic undercut etch is performed at a sample holder temperature of 250 $^{\circ}$C, completing the total undercut in only 64 minutes. The fabricated waveguides are longitudinally aligned along the [110] direction.

\begin{figure}[h!]
	\includegraphics[width=0.8\textwidth]{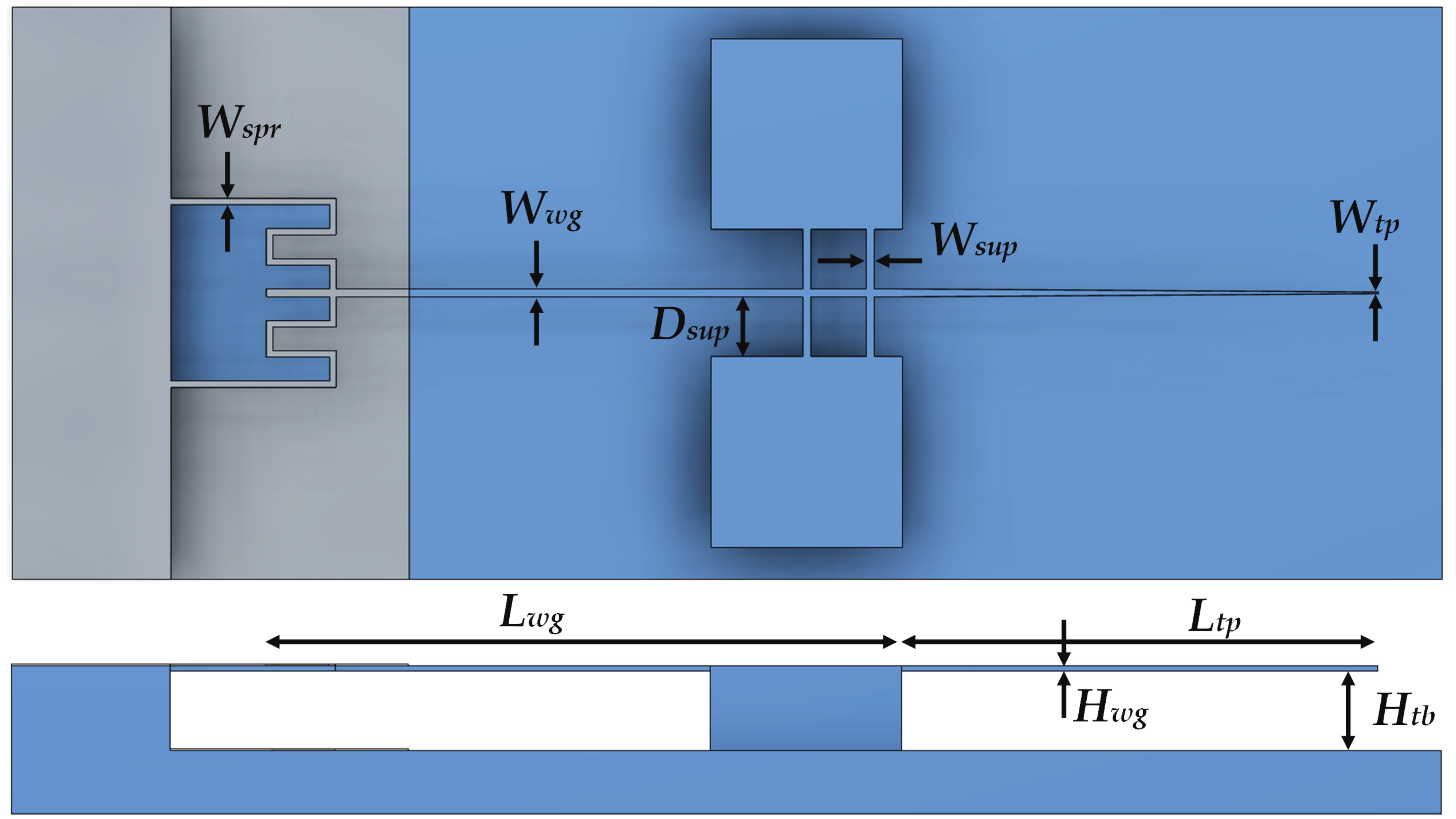}
	\caption{Schematic overview of the device and the relevant parameters of the geometry.} \label{fig:device geometry}
\end{figure}

Finally, we fabricate thin-film electrodes onto the diamond nanostructures via lift-off. A bi-layer positive-tone electron-beam resist stack consisting of methyl methacrylate (MMA) and polymethyl methacrylate (PMMA) is used to facilitate the lift-off of the electrodes. The electrode material stack consists of a 6 nm thick titanium adhesion layer and 10 nm thick niobium film, both deposited via electron-beam evaporation in a single evaporation run. These films are deposited at the lowest rate possible with the used evaporator tool to minimize stresses in the films, which could lead to detachment of the electrodes during lift-off. The lift-off process of the electrodes is performed in a heated bath of acetone. A schematic overview of the device and the relevant parameters are given in Fig.~\ref{fig:device geometry} and Tab.~\ref{tab:geometry}.

\begin{table}[]
\begin{tabular}{|l|l|}
\hline
\textbf{Parameter}        & \textbf{Value}                 \\ \hline
$W_{\text{spr}}$ & 200 nm                \\ \hline
$W_{\text{wg}}$  & 250 nm                \\ \hline
$W_{\text{sup}}$ & 250 nm                \\ \hline
$W_{\text{tp}}$  & 50 nm                 \\ \hline
$D_{\text{sup}}$ & 2 $\upmu$m             \\ \hline
$L_{\text{wg}}$  & 20 $\upmu$m            \\ \hline
$L_{\text{tp}}$  & 15 $\upmu$m            \\ \hline
$H_{\text{tb}}$  & $\approx$ 2.5 $\upmu$m \\ \hline
$H_{\text{wg}}$  & 160 nm                \\ \hline
\end{tabular}
\caption {Parameters of the device geometry.}
\label{tab:geometry}
\end{table}
\section{Experimental set-up}
\begin{figure}
	\includegraphics[width=\textwidth]{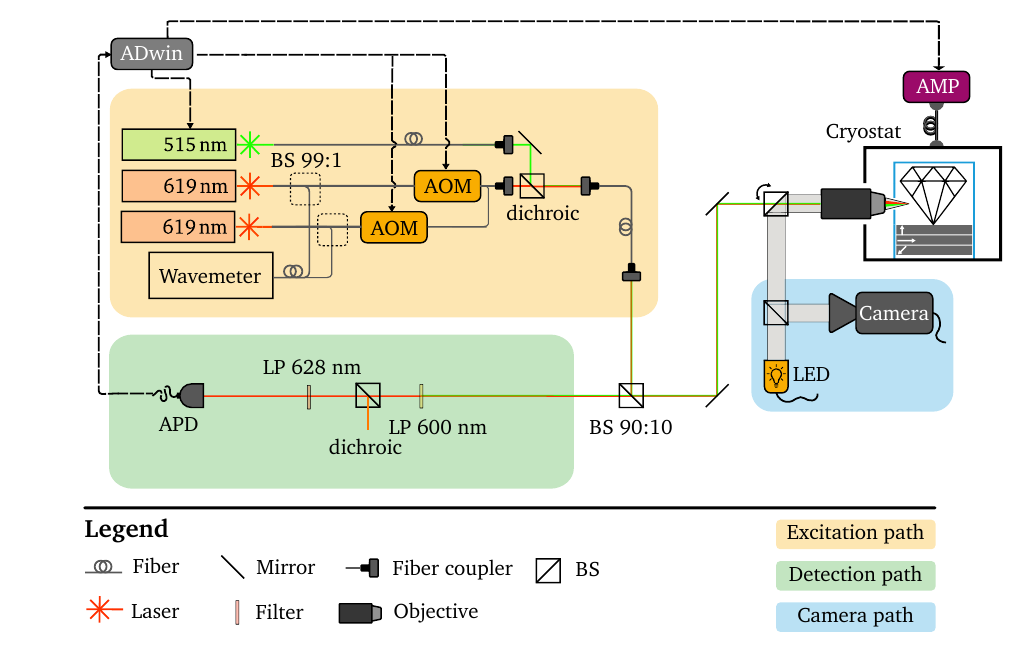}
	\caption{Experimental confocal set-up} \label{fig:confocal_setup}
\end{figure}

All measurement data in this work has been taken in a closed-cycle attoDry800 cryostat at a baseplate temperature of \unit[4]{K}. The diamond sample is glued on a PCB that is mounted on attocube positioners (ANPz102, ANPx101 (2x), ANSxyz100). The positioner stack and the PCB are enclosed by an aluminum heat shield to block radiation from the objective. All resonant excitation (\unit[619]{nm}) is done by using a Toptica TA-SHG Pro and a Toptica TA-SHG. The wavelength of the excitation light is stabilized by feedback through a wavemeter (HighFinesse WS-6). All off-resonant \unit[515]{nm} excitation is done by a Hubner Photonics Cobolt 06-MLD. The resonant optical pulses are created using in-fiber acousto-optic modulators (Gooch and Housego) controlled by a microcontroller (Jaeger ADwin Pro II). The laser paths of the different lasers are combined in a single path and focussed on the diamond region of interest. To obtain only the PSB light, the resonant and off-resonant laser photons are filtered by a long-pass (Thorlabs, FELH0600), dichroic beamsplitter (Semrock, FF625-SDi01), and 2 tunable \unit[628]{nm} long-pass (Semrock, TLP01-628) filters in the detection path. We use an APD for photon detection which is connected to the microcontroller for photon counting. We use the DAC channel of the microcontroller and an in-house build voltage amplifier (x10, \unit[2]{$\upmu$s} ramp time) to deliver the voltage pulses for strain engineering. The experimental setup is controlled by a PC and the Python 3 framework QMI 0.37~\cite{raa_qmi_2023}. A schematic overview of the set-up is depicted in Fig.~\ref{fig:confocal_setup}.

\section{Voltage pulse calibration}
We observe that for large applied voltages leakage current through the system caused significant resistive heating. We note that on this chip 48 strain-tune devices are biased in parallel so that a leakage current at one of the devices affects the operation of all of them. This effect can be mitigated in future chips by adding individual voltage biasing to each device. This local increased temperature lowers the optical resonant frequency of SnV$^-$ and broadens the optical linewidth, which is detrimental for quantum applications that rely on indistinguishable photons. By pulsing the applied bias voltage to induce local strain, the effect of the heat can be minimized. We calibrate the minimal required cool-down time after the voltage pulse by sweeping a waiting time after the voltage pulse during PLE scans. For cool-down times longer than ~\unit[1500]{$\upmu$s}, no significant change in the resonant frequency is obtained. Secondly, we calibrate the voltage pulse duration to eliminate heat during the voltage pulse, by sweeping the pulse duration. We evaluate the resonant frequency and FWHM by fitting the collected PSB photons during the pulse in \unit[1]{$\upmu$s} time bins, for different durations of the pulse. For pulses of \unit[50]{$\upmu$s}, we observe no change in resonant frequency and FWHM during the duration of the pulse. In addition, we observed no hysteresis in the voltage when it was applied in a pulsed way.

\section{FEM simulations}
The FEM simulations in the main text are performed using the simulation software Ansys~\cite{noauthor_ansys_nodate}. The FEM simulation solves for an equilibrium of the restoring forces of the dielectric and the electrostatic forces. The electrostatic field between the top and bottom electrode is \unit[0.3]{MV/cm} at a voltage of \unit[70]{V}. The model outputs the components of the strain tensor in the geometry of our device. The effective thermal expansion coefficients for diamond and niobium are determined from the data in~\cite{dias_uranium-zirconium_2015,jacobson_thermal_2019}, by fitting the data and then integrating the fit from room temperature, \unit[293]{K}, to the temperature at which the measurements were performed, \unit[4]{K}.

\begin{table}[]
\begin{tabular}{|l|l|l|}
\hline
\textbf{Material paramters  }                & \textbf{Diamond}         & \textbf{Niobium}        \\ \hline
Young's modulus (MPa)               & 1050 x 10$^3$~\cite{klein_youngs_1993}  & 105 x 10$^3$      \cite{shabalin_ultra-high_2014}   \\ \hline
Poisson's ratio                     & 0.2  \cite{klein_youngs_1993}             & 0.35  \cite{shabalin_ultra-high_2014}           \\ \hline
Electrical resistivity ($\Omega$m)    & -               & 1.52 x 10$^{-7}$ \cite{shabalin_ultra-high_2014}  \\ \hline
Thermal expansion coefficient (1/K) & 0.282 x 10$^{-6}$  \cite{jacobson_thermal_2019} & 5.06 x 10$^{-6}$  \cite{dias_uranium-zirconium_2015} \\ \hline
Dielectric constant                 & 5.68            & -              \\ \hline
\end{tabular}
\caption {Material properties used for the FEM simulations}
\end{table}

\end{document}